# Finite Element Model Updating-based Load Rating of Bridges with Incomplete As-Built Information

[1]Mehrdad S. Dizaji, [2]Devin K. Harris, [3]Mohamad Alipour, [4]Abdou Ndong, Osman Ozbulut and Abdollah Bagheri

[1] Postdoctoral Research Associate, Engineering Systems and Environment, University of Virginia, U.S.

[2] Professor, Engineering Systems and Environment, University of Virginia, U.S.

[3] Research Assistant Professor, Civil Engineering, University of Illinois at Urbana Champaign, U.S.

[4] Graduate Student, Engineering Systems and Environment, University of Virginia, U.S.

[5]Associate Professor, Engineering Systems and Environment, University of Virginia, U.S.

[6] Research Associate, Civil Engineering, University of Maryland-College Park, U.S.

**Abstract**

Load rating is the engineering process of determining the safe load-carrying capacity of an existing bridge structure often through analysis of its load-bearing members and their cross sections. However, when structural drawings and details of the structure are missing or insufficient for such calculations, alternative solutions must be used to infer the capacity. For such cases, the American Association of State Highway and Transportation Officials Manual for Bridge Evaluation (AASHTO MBE) recommends the use of estimation by engineering judgment or the experimentally-based proof testing. However, the former approach is known to be subjective while the latter can be prohibitively expensive or unsafe. This study proposes a rational and consistent engineering solution for capacity inference based on structural identification (St-Id). The proposed approach uses finite element model updating (FEMU) to estimate the unknown characteristics of structures for use in an analytical load rating, which requires the development of an initial model developed in ABAQUS that can be updated based on experimental test data. ABAQUS allowed for the development of an interface with MATLAB, which facilitated the integration of an automatic iterative parameter optimization algorithm. The optimization algorithm developed in this investigation incorporated the features of a genetic algorithm (GA) and a gradient-based scheme to iterate on the unknown parameters. The proposed approach was evaluated on four in-service highway bridges including two concrete slab and two T-beam structures in varying deterioration condition states, which had sufficient plans available but were treated as having varying degrees of unknown details.

[1] Corresponding Author: Postdoctoral Research Associate, Engineering Systems and Environment, University of Virginia, U.S

The results illustrated that the finite element (FE) model updating approach generated load ratings that were within 0-17% of the target load ratings while alleviating the main challenges of the existing alternatives. Sensitivity analysis of the proposed method also showed that instrumenting bridges with a limited number of sensors is sufficient for successful implementation of the developed methods.

## 1. Introduction

Load rating is the process of determining a bridge's safe load-carrying capacity; however, when plans and details are insufficient to determine the nominal capacity, other methods must be used to infer the remaining capacity. In such cases, the AASHTO MBE [1] recommends two methods for bridge evaluation: the commonly used but subjective estimation via engineering judgment, and the experimentally-based proof testing. These methods, however, have significant limitations. Estimation via engineering judgment relies on the expertise and experience-driven judgment of a qualified engineer, but is not typically based on measurements of physical phenomena, which creates an unknown degree of variability and uncertainty resulting either in unconservative estimates or unnecessary restrictions of traffic and commerce in overly conservative estimates. On the contrary, proof testing can cause damage during testing, tends to be expensive, and cannot be extrapolated to future performance [1, 2].

To address these challenges, researchers have proposed a number of approaches to tackle this problem over the past few years. Shenton et al. [3] and Huang and Shenton [4] investigated a method for load rating of Reinforced Concrete (RC) bridges without as-built information. The method used strain or displacement data to estimate the unknown area of reinforcing steel in a bridge. This estimated reinforcing steel area was then used to determine the bridge’s capacity based on a sectional analysis procedure. To utilize the proposed approach, strain sensors needed to be mounted on the top and bottom surface of a deck to quantify the internal strain distribution of the bridge’s cross-section. This approach also required knowledge of the relevant mechanical properties for the selected materials. Subedi [5] used non-destructive technologies including a concrete rebound (Schmidt) hammer and cover meter to estimate concrete strength and rebar details, respectively and compared the results with existing plans for a flat slab bridge. A finite element model of the structure was then built and the bridge model was loaded with the state legal truck. The load was increased such that the model reached AASHTO's maximum serviceability deflection. The ratio of the corresponding allowable load to the original design load was then calculated designed as a rating factor. It should be noted that the method did not include the effects of field factors or bridge condition in load rating, and the use of deflection as the limit state does not correspond with the actual allowable capacity for load rating calculation. Aguilar et al. [6] presented a four-step load rating procedure for prestressed concrete bridges without plans using proof load test results. Their proposed procedure included estimating the number and eccentricity of strands using Magnel diagrams and typical details at the time of construction. Material properties were assumed based on age using the nominal values recommended by the MBE [1]. A rebar detection system was used to detect the location and size of the reinforcement and concrete cover as a verification of the estimates. A diagnostic load test was then conducted to determine the critical transverse truck path and to estimate load effects under the trucks. Finally, a proof load test was performed with a target proof load based on the MBE and the results were used to determine final load ratings. Taylor et al. [7] presented a method to load rate the bridges with small to medium simple span bridges in Larimer County, CO that are currently load rated solely based on visual inspections and estimation by engineering judgment. Due to the absence of the structural drawings, a basic structural analysis was performed using a program developed for Colorado Department of Transportation in order to determine the capacities of the bridges. Briones [8] proposed a general procedure for load rating bridges without plans that included bridge characterization, bridge database, field survey and inspection, and bridge load rating. The bridge characterization focused on the identification of critical bridge information needed to conduct the load rating and evaluation. The bridge database provided guidelines and recommendations for obtaining the unknown information discerned from the bridge characterization. A field survey was used to supplement the unknown bridge information through collecting field measurements. The load rating of bridges without plans was successfully completed by following the developed general procedure for two case study bridges without plans.

Field tests results of old bridges show that there is a considerable reserve capacity in terms of strength in most of the bridges that is not justified by the rating procedure within the standards of AASHTO which classify them as structurally deficient. Azizinamini et al. [9] outlined an experiment of aged reinforced concrete bridges both at service and ultimate levels to rate them more realistically. To accomplish these objectives, six concrete slab bridges were tested under the selected weights of truck loads so that the bridge responses would be confined to the elastic regime (service load tests). Also, a five-span reinforced concrete bridge built in 1938 was tested destructively, which was performed by applying loads that simulated two trucks side by side on the structure. Experimental test results show that the reinforced slab bridges have much higher strengths than indicated by AASHTO rating procedures. Chajes et al. [10] conducted an experimental load rating of a posted, three-span, slab, and steel-girder-and-slab bridge. Each span of the bridge consisted of a cross-section of nine non-composite steel girders, with the outer girders spaced 1.37 m apart and the interior girders spaced 1.52 m on center. They conducted a load diagnostic test and found that the girders act compositely with the concrete deck and a high restraint observed at the supports. Along with the diagnostic test, a predetermined load was placed at several different locations along the bridge and the bridge response was measured. The measured response was then used to develop a numerical model of the bridge. This numerical model was employed to determine the maximum allowable load by applying the load incrementally until a target load was attained or a predetermined limited state was exceeded. The results indicated that the bridge's load-carrying capacity may be substantially higher than the current load levels indicated and suggested that the posting levels on the bridge may be unnecessary. Cai and Shahawy [11] conducted a load test on six prestressed concrete bridges with different geometric characteristics to evaluate analytical methodologies for load rating, which were shown to be unreliable [12]. The main objective was to compare the results from the measurements obtained from their study with AASHTO codes specifications and with those ratings predicted using FE analysis. The comparison showed a notable difference between the analytical and experimental due to the effects of several factors. However, to examine these effects, the authors included some field factors in their FE models which in turn had a larger effect on the maximum strain than on the load distribution factor. Parametric studies on the effects of the components of the bridges were also carried out in these studies to assess how the distribution and maximum strain were affected.

Turer and Shahrooz [13] presented an investigation of the use of 2D grid models for field-calibrated model based load rating of concrete deck on steel stringer bridges. The main hypothesis in this work was that 2D grid models are an efficient tool for modeling concrete deck on steel stringer bridges which will then be used for model calibration against bridge tests and the calibrated models can be finally used for load rating. The 2D grid model used in this paper employed linear elastic beam elements configured as a grid simulating the entire superstructure and the deck. As to their model calibration, the researchers used a code written in MATLAB based on an automatic updating algorithm which performs structural analysis and objective function optimization. As to the response variable used for updating, they used both modal data (frequencies, modal assurance criteria (MAC) and order of modes) and static deformations (BGCI-Bridge Girder Condition Index). The proposed method was then applied on an actual three-span four-lane concrete deck on a steel stringer bridge built in 1968. Load ratings were calculated using the allowable stress rating (ASR) and the load factor rating (LFR). The authors concluded that the updating scheme had been efficient and successful, that 2D grid models provided results close to 3D models and that transverse members were the critical and controlling members in the system's load rating.

The objective of this study was to develop rational engineering approaches for load rating structures within the Virginia Department of Transportation (VDOT) inventory for which limited as-built information is available. The initial phase of the investigation focused on categorizing the VDOT inventory to determine the types of structures that are likely to be missing information necessary for an analytical load rating, which were short span reinforced concrete slab or T-beam designs. Subsequent phases emphasized two main approaches to load rating: (i) structural identification frameworks based on finite element model updating; and (ii) leveraged vibration response characterization. Both approaches emphasized estimating unknown characteristics of these types of structures for use in a traditional analytical load rating. These unknown parameters include modulus of elasticity and strength of concrete as well as cross-sectional area of steel reinforcement. These estimates can ultimately be used to provide a rational estimate of load ratings.

## 2. Proposed Methodology

The research team explored a series of experimental and numerical approaches on four bridges within VDOT's inventory. The described approaches were initially developed and evaluated on one bridge, then refined and extended to an additional three bridges. All of the bridges evaluated in this study had sufficient plans and other information necessary to perform a traditional analytical load rating, but served as representative structures similar to those bridges without plans that were derived from the characterization portion of this study. With the consent of the project's technical review panel (TRP), the selections included structures in "good" and "fair" condition to allow for condition characteristics to be included in the assessment methods. The other qualification was the suitability for field testing (that is, proximity, accessibility, and low average daily traffic). This strategy allowed for a ground truth reference of the proposed methodologies, which were developed under the assumption that the necessary information for the conventional analyses was unavailable.

Building upon the principles outlined in the MBE, analytical and experimental approaches can be combined to create an effective and realistic rating method that aims to leverage the strengths of each one of these approaches while reducing the operational difficulties of load testing. This method relies on the application of structural identification (St-ID) for FE model updating based on live load testing, vibration testing, or both. The term Structural Identification (St-ID) is defined as the process of creating/updating a structural model based on experimental observations/data [14-18]. The St-ID framework aims to bridge the gap between model approximation and the real system behavior through improved simulations. One of the subcomponents of St-ID is FE model updating (FEMU), which as noted requires the development of an initial model that can be updated based on experimental test data. An updated model is expected to reflect the measured data better than the initial model as a result of refinements to model uncertainties (e.g. boundary/loading conditions and constitutive properties).

In this investigation, initial models were developed in ABAQUS, a robust commercially available finite element software package [19]; however, comparable FEA models could be developed in other packages. In this study, ABAQUS allowed for the development of an interface with MATLAB [20], which facilitated the integration of an automatic iterative parameter optimization algorithm. The optimization algorithm systematically searches a parameter space by revising input variables to achieve a solution match, in this work minimizing the difference between the model and experimental results by iterating on uncertain parameters. The optimization algorithm developed in this investigation incorporated the features of a genetic algorithm and a gradient-based scheme to iterate on the unknown parameters [21]. A genetic algorithm (GA) is a global optimization method and provides an effective means to solve for unknowns whenever there are more than two unknown parameters in the problem. Therefore, using a GA, the chance of getting trapped in local minimums decreases dramatically. However, the proposed GA has a relatively high computational cost, which results in a long solution time. In order to reduce the computational cost, this method can be combined with a gradient-based algorithm. While other optimization algorithms could be used, this approach was selected because of the efficiency in localizing a global minimum amongst a large dataset [22].

In this investigation, area of steel and material properties were selected as key unknown parameters and boundary restraints as secondry within the St-ID scheme as these values have time-dependent and condition-driven characteristics. In addition to identifying unknown boundary properties, the updating process allowed for identification of some of the key parameters that are needed to determine load rating, including the area of steel, $A_s$, the elastic modulus of concrete, $E_c$, and the concrete compressive strength, $f'_c$. A schematic of the overall model updating framework used is shown in Figure 1, and can be described as a process that minimizes the difference between the response of the bridge obtained in an FE model and those obtained from experimental testing.

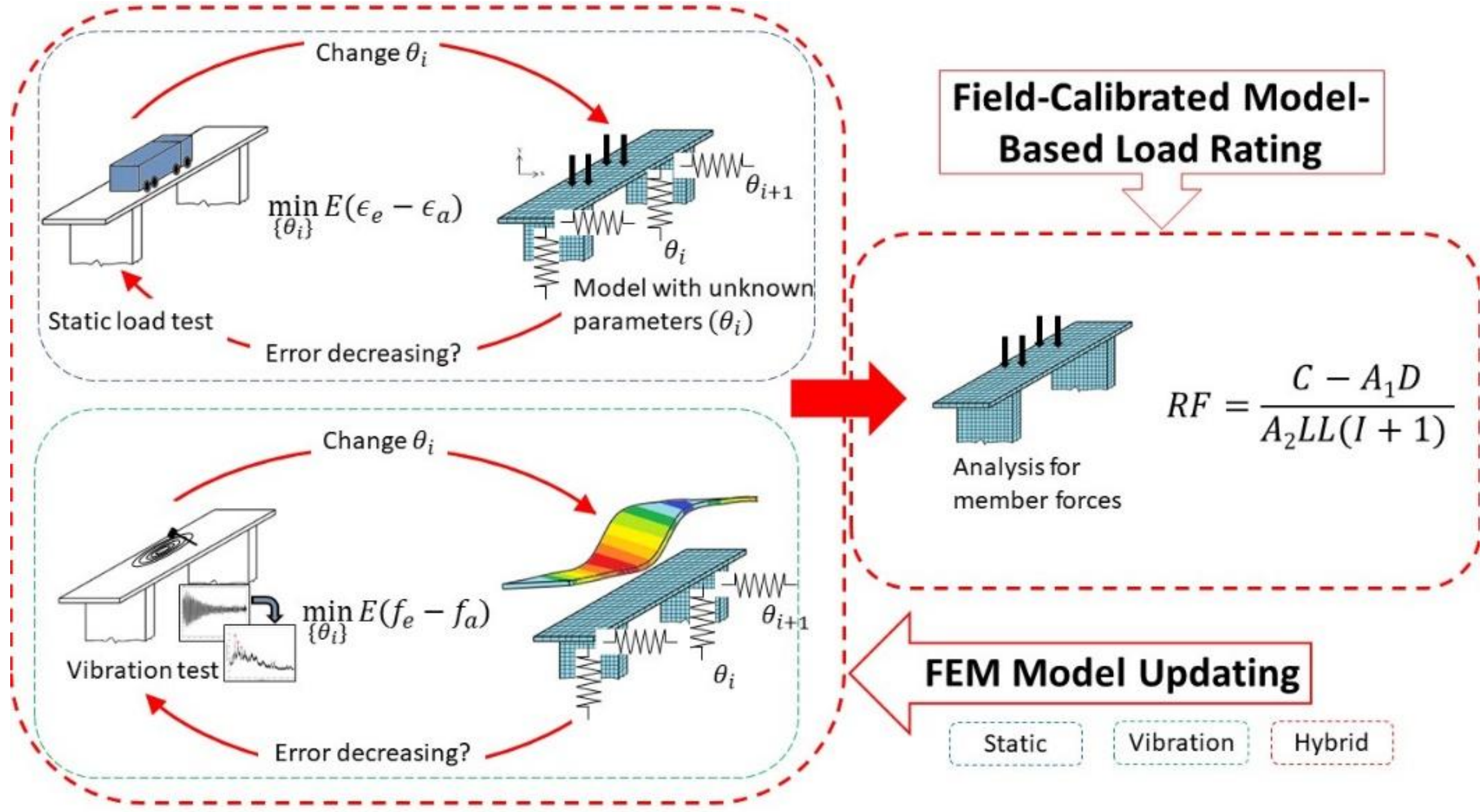

Figure 1. Schematic of Load Rating Using FEMU Based on Field Test Data

In order to determine the unknown structural parameters, optimization was performed to minimize the difference between calculated data from a numerical model and a set of experimental data. This research developed an FE model with parameterized geometry and boundary conditions defined to allow the model to replicate the real conditions and well-known geometry of the test bridges as closely as possible once updated. The results from the FE model were compared with experimental data collected from sensors on a given bridge and the objective functions were evaluated for convergence [23, 24]. The parameters were then iteratively updated in an optimization process until convergence. To compare the effect of the type of sensing data used for model updating, three different FE model updating scenarios were evaluated and are described in the following sub-sections:1) Discrete Static FE Model Updating (FEMU-S), 2) Discrete Dynamic FE Model Updating (FEMU-D), 3) Discrete Hybrid FE Model Updating (FEMU-H). The sensor types used and their corresponding generalized objective function are summarized in Table 1 and further details are given below. However, it should be noted that in all of the FEMU scenarios, the optimization processes aimed to solve for the critical unknowns needed to perform the load rating, which for a concrete structure generally consisted of area, location, and distribution of the reinforcing steel, as well as the compressive strength and elastic modulus of the concrete.

Table 1. General Form of Objective Functions for FEMU Scenarios

| **FEMU Type** | **Objective Function** |
|---|---|
| FEMU-S | $F(E_c,\ A_s, K_{\theta x},\ K_{\theta z}) = \sum_i \frac{\lvert \varepsilon_i^{Measure} - \varepsilon_i^{FEM} \rvert}{\varepsilon_i^{Measure}} + \sum_i \frac{\lvert \delta_i^{Measure} - \delta_i^{FEM} \rvert}{\delta_i^{Measure}} + \sum_i \frac{\lvert \theta_i^{Measure} - \theta_i^{FEM} \rvert}{\theta_i^{Measure}}$ |
| FEMU-D | $F(E_c,\ A_s, K_{\theta x},\ K_{\theta z}) = \sum_i \frac{\lvert \omega_i^{Measure} - \omega_i^{FEM} \rvert}{\omega_i^{Measure}}$ |
| FEMU-H | $F(E_c,\ A_s, K_{\theta x},\ K_{\theta z}) = \sum_i \frac{\lvert \varepsilon_i^{Measure} - \varepsilon_i^{FEM} \rvert}{\varepsilon_i^{Measure}} + \sum_i \frac{\lvert \delta_i^{Measure} - \delta_i^{FEM} \rvert}{\delta_i^{Measure}} + \sum_i \frac{\lvert \theta_i^{Measure} - \theta_i^{FEM} \rvert}{\theta_i^{Measure}} + \sum_i \frac{\lvert \omega_i^{Measure} - \omega_i^{FEM} \rvert}{\omega_i^{Measure}}$ |

In Table 1, $E_c$ and $A_s$ were selected as unknown parameters in the optimization scheme along with $K_{\theta x},\ K_{\theta z}$ which are restraint stiffness of the rotational springs at the support locations. $\varepsilon_i^{Measure}$ is the longitudinal measured strain obtained from strain gage sensors installed on the bridge in the location shown in Figure 2 with index *i* and $\varepsilon_i^{FEM}$ is the longitudinal numerical strain obtained from the FE model of the bridge in the location shown with index *i*. $\delta_i^{Measure}$ is the measured vertical displacement obtained from string potentiometer sensors installed on the bridge in the location shown with index $i$ and $\delta_i^{FEM}$ is the numerical vertical displacement obtained from FE model of the bridge in the location shown with index *i*. $\theta_i^{Measure}$ is the measured rotation obtained from tiltmeter sensors installed on the bridge in the location shown with index $i$ and $\theta_i^{FEM}$ is the numerical rotation obtained from FE model of the bridge in the location shown with index *i*. $\omega_i^{Measure}$ is the derived modal frequencies obtained from accelerometer sensors installed on the bridge in the location shown with index *i* and $\omega_i^{FEM}$ is the numerical modal frequencies obtained from FE model of the bridge in the location shown with index *i*. Some of these objective function's parameters are shown in Figure 2 schematically.

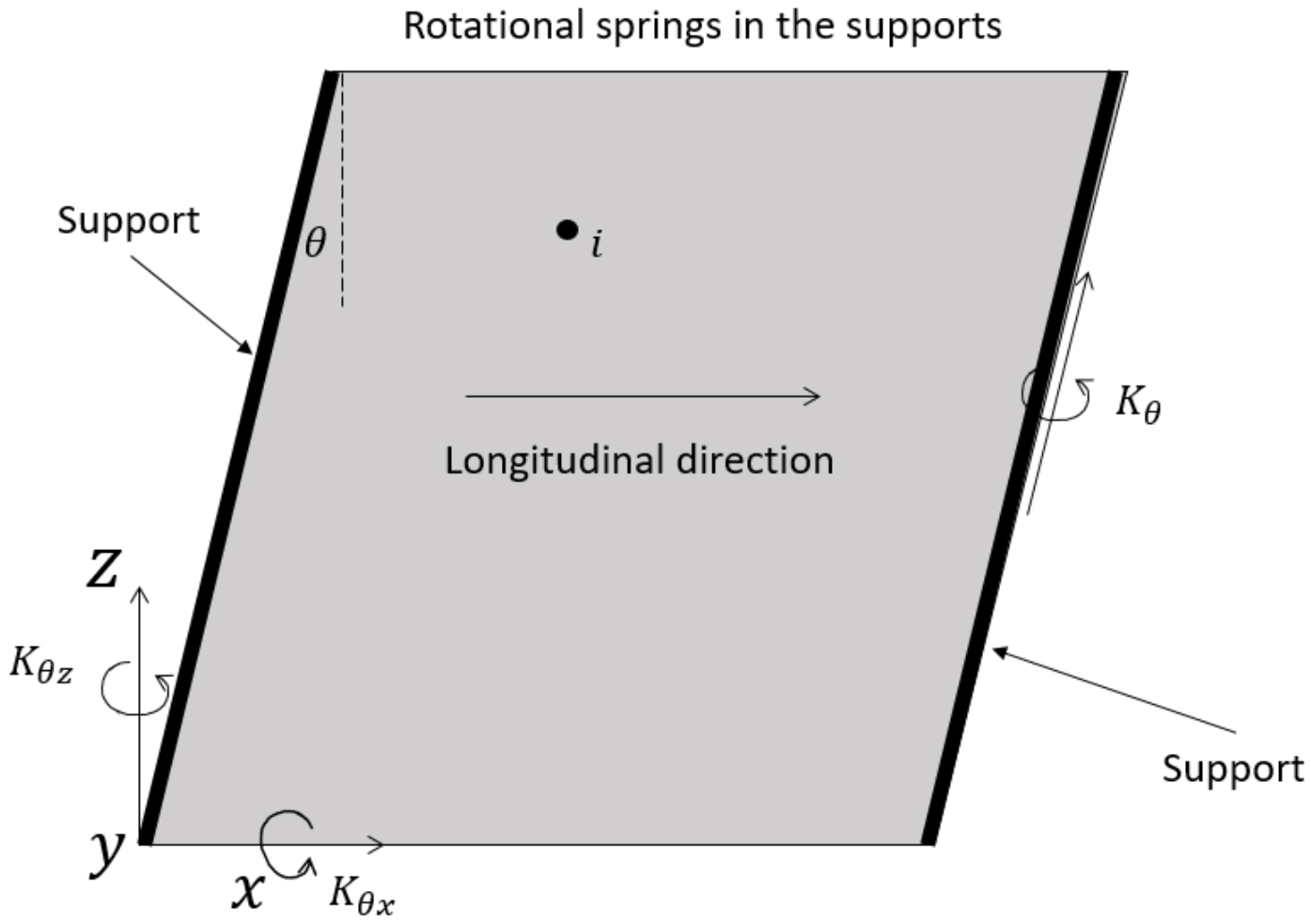

Figure 2 - Parameters of Objective Function Shown Schematically

**2.1. Finite Element Model Updating Methods**

For FEMU-S only quasi-static sensors were used in model updating. This type of measurement would be typical of a traditional live load test. Time series data was post-processed and the sensor data for the critical loading configuration (peak measured response) was used in the updating processes. For FEMU-D the first three identified natural frequencies of the bridge (derived from vibration testing) were used in model updating. This type of measurement would be typical of traditional ambient vibration or an impact excitation test. This scenario was similar to the previous method, except that the FE model was updated using an array of sensors used to record the vibration response of a bridge and extract the modal properties of the structure. In this method, an output-only modal identification technique called the Enhanced Frequency Domain Decomposition (EFDD) was used to obtain modal characteristics of bridge structures (i.e. first three natural frequencies). In this method, the single degree of freedom (SDOF) power density functions are transferred back into the time domain, and the natural frequencies are obtained by calculating the number of zero-crossings as a function of time. In addition, the damping ratio can be estimated from the logarithmic envelope of the corresponding SDOF correlation function using the logarithmic decrement method. The EFDD technique allows for the extraction of the natural frequency and damping characteristics of a particular mode by processing the collected time domain acceleration data in frequency domain. Additional details on the EFDD technique can be found in Brincker [25]. In this study, the results for the system identification techniques were obtained from the ARTeMIS Modal Pro software [26]. Similar to the FEMU-S approach, the experimental data was used to update a model and estimate the rating factor. For FEMU-H both mechanical and accelerometer sensors were used in model updating. This method designated as the hybrid approach, combines the FEMU-S and FEMU-D approaches to simultaneously include both global and local behavior characteristics of the bridge. Time series data was post-processed and the sensor data for the critical loading configuration (peak measured response) for the mechanical response along with the first three natural frequencies derived from the accelerometers using the EFDD were used in the updating processes. Similar to the FEMU-S and FEMU-D approaches, the experimental data was used to update a model and estimate the rating factor (Figure 3):

1. An approximate FE model of a bridge was first developed. This model was based on readily observable geometric parameters, reasonable estimates of unknown material properties, and assumptions regarding internal structural parameters, such as cross-sectional area and location of the steel reinforcement.

2. The model was then updated based on measured data obtained from sensors. Measured responses such as vibration, strain, deflection, and rotation were used as inputs to update the model of the bridge.

3. An inverse method based on an optimization process was then applied to find unknown structural properties of the structure such as stiffness and boundary conditions. The Young's Modulus ($E_c$) of the concrete and area of steel ($A_s$) of rebar were selected as unknown parameters in the optimization scheme along with restraint stiffness at the support locations.

4. Based on the identified concrete elastic modulus ($E_c$), area of steel of rebar and an assumption of dynamic amplification ($IM$), the capacity of the structure ($C$) was calculated

5. Finally, the rating factor ($RF$) was obtained using Equation from Table 1.

An illustration of each of the FE models is shown in Figure 4, with representations of the model details and internal features included in the model development.

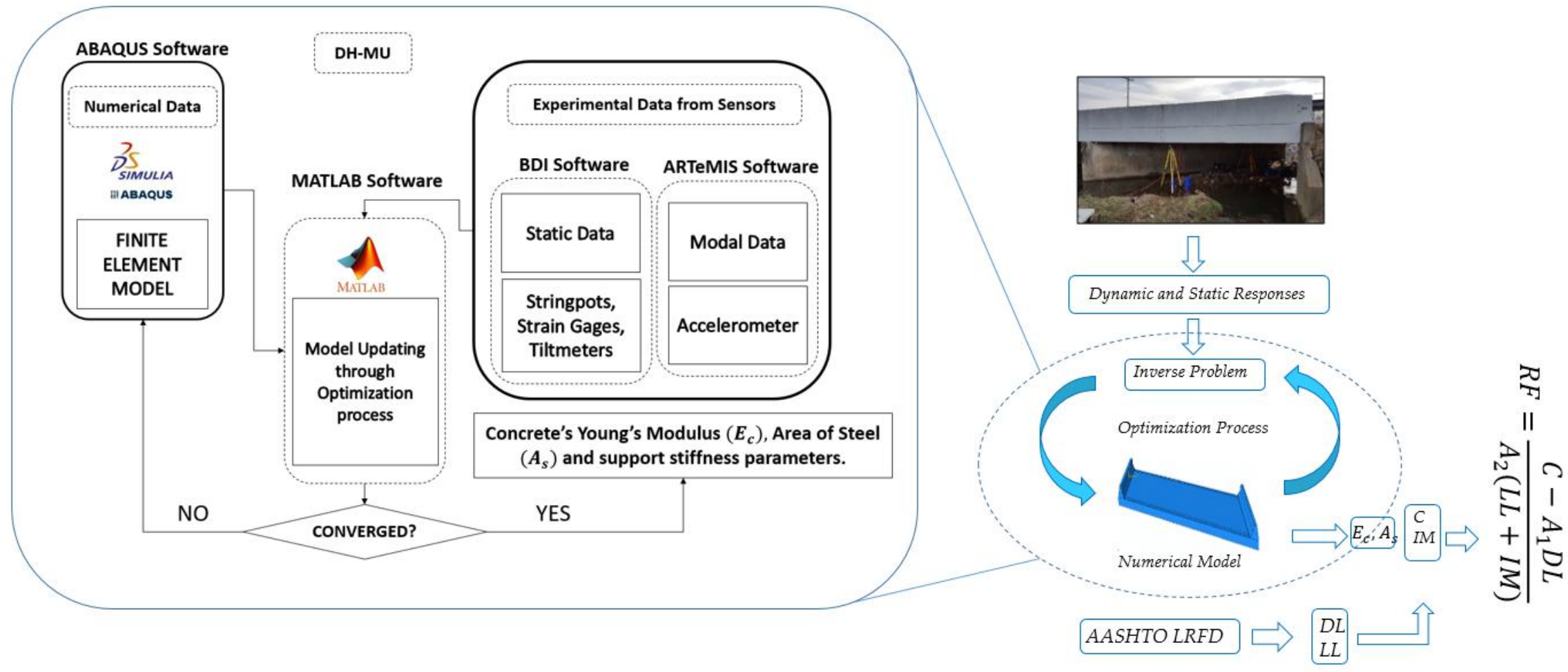

Figure 3. Description of the proposed methods for FEMU-S, FEMU-D and FEMU-H

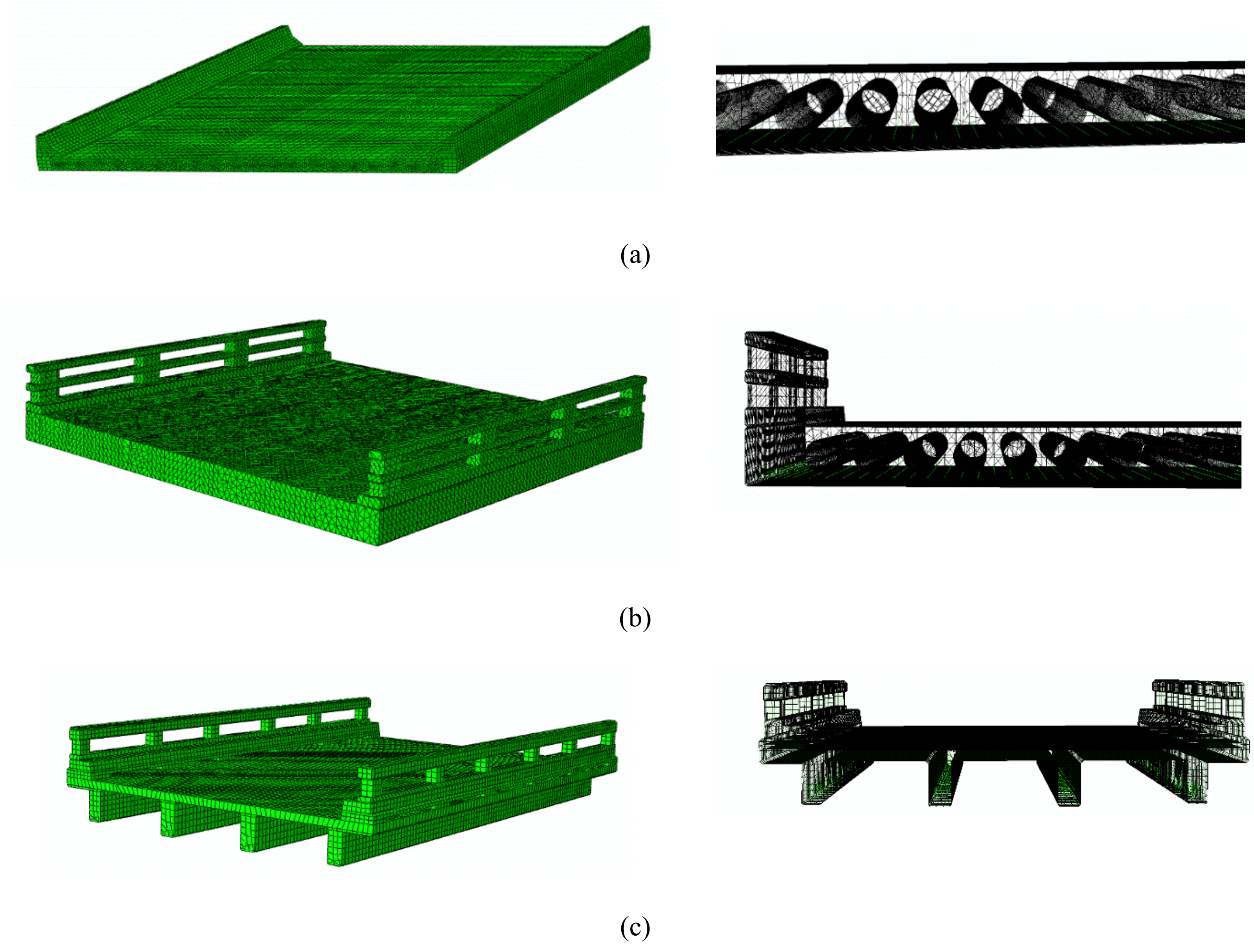

(a)

(b)

(c)

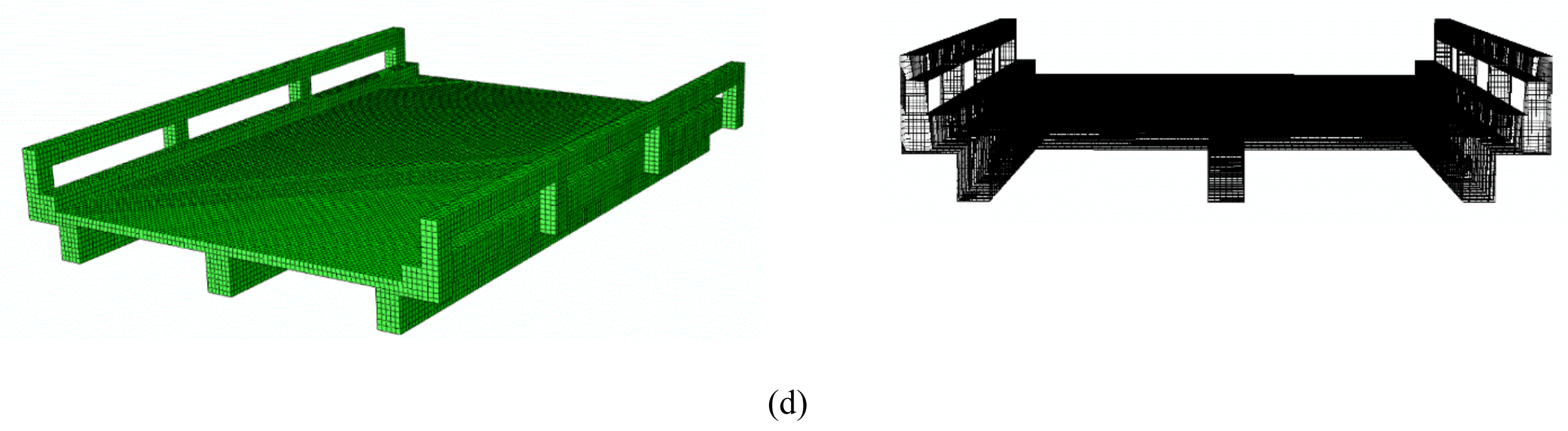

(d)

Figure 4 – Preliminary Solid FE Model and Wireframe Representations to Illustrate Internal Geometry Assumptions (a) War Branch Bridge; (b) Smacks Creek Bridge; (c) Flat Creek Bridge; and (d) Brattons Creek Bridge

## 3. Characterization of VDOT Inventory of Bridges with Limited Information

To understand the scale and scope of the problem, analysis of the inventory of in-service highway bridges within the Commonwealth of Virginia (12,925 structures) determined that approximately 7% of these structures (933 structures) did not have plans. Within this pool of 933 structures, the vast majority were reinforced concrete structures (Figure 5a), with the primary design configurations being slab, arch-deck, and T-beam bridges (Figure 5b). This study investigates sample structures from the concrete slab and T-beam classes, which combined, represent 63% of the population of bridges without plans.

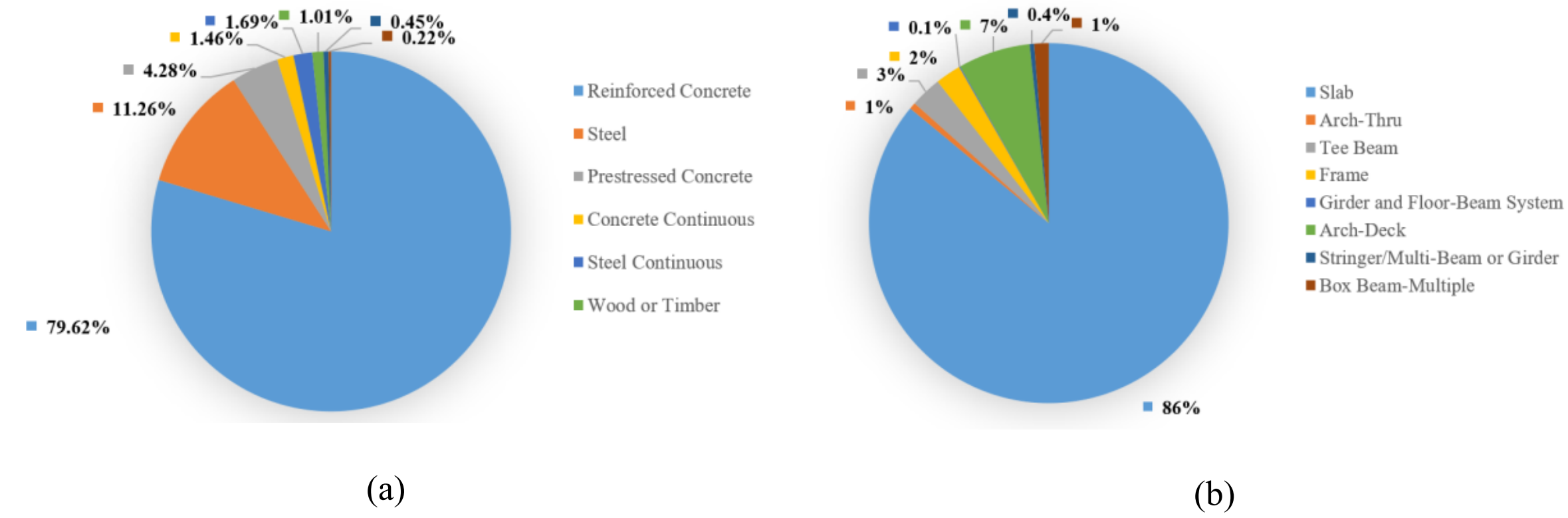

(a) (b)

Figure 5. Distribution of VDOT Bridges without Plans by (a) Primary Structural Material and (b) Bridges Built with Reinforced Concrete Material

To evaluate the performance of the proposed methods for load rating of bridges with unknown or insufficient details, four test bridges were selected including two slab bridges and two T-beam bridges. The bridges selected were in good and fair condition, one in each condition state for each bridge category. While plans were available for each bridge and provided a basis for comparison of the results derived from each method, their information were not used throughout the process to mimick the target case of bridges with missing information. A summary of the selected bridges is as follows (additional details on the selected bridges are available from VDOT):

*War Branch Bridge* (Figure 6a): The superstructure is comprised of two 32-ft long, simply-supported reinforced concrete slabs that are 21 inches thick and have a 45° skew. The deck has 12-inch diameter voids oriented in the direction of traffic, spaced 18 inches apart. Built in 1976, the most recent inspection report described the bridge to be in “good” condition, with a deck/superstructure condition rating of 7.

*Smacks Creek Bridge* (Figure 6b): The superstructure is comprised of two 32-ft long, simply-supported reinforced concrete slabs that are 21 inches thick and have a 15° skew. The deck has 12-inch diameter voids oriented in the direction of traffic, spaced 18 inches apart. Built in 1965, the most recent inspection described the overall condition of the bridge to be in “fair” condition, with a deck/superstructure condition rating of 5. The concrete slabs had cracking, delamination, efflorescence, rust staining, pop-outs and scaling. The pourable joint sealer was brittle and deteriorated. The substructure had cracking and delamination with efflorescence and minor spalling.

*Flat Creek Bridge* (Figure 6c): There are five simple spans, each 42.5 ft long, for a total length of 212 ft. Each span is 24 ft wide and consists of three longitudinal T-beams. Each exterior T-beam has a vertical stem with a width of 14 in and a depth of 32 in. Each interior T-beam has a vertical rectangular stem with a width of 16 in and a depth of 32 in, and a wide top flange of 7.5 in thick. The wide top flange is the transversely reinforced deck slab. Built in 1950, the most recent inspection described overall condition of the bridge to be in “good” condition, with a deck/superstructure condition rating of 7. The structure had isolated areas of minor spalling and scale.

*Brattons Creek Bridge* (Figure 6d): The structure is comprised of three simple spans of the same length, 32 ft, with a total length of 98 ft - 2in, a width of 23 ft - 8 in, and consists of three longitudinal T-beams. Each exterior T-beam has a vertical stem with a width of 14 in and a depth of 24 in. The interior T-beam has a vertical rectangular stem with a width of 16 in and a depth of 24 in, and a wide top flange of 8 in thick. The top flange is the transversely reinforced deck slab and the riding surface for the traffic. Built in 1953, the most recent inspection described the overall condition of the bridge to be in "fair" condition, while the deck condition rating was 6. The report noted spalls, cracks and delamination in the superstructure. The substructure exhibited scaling, cracking and signs of channel drift.

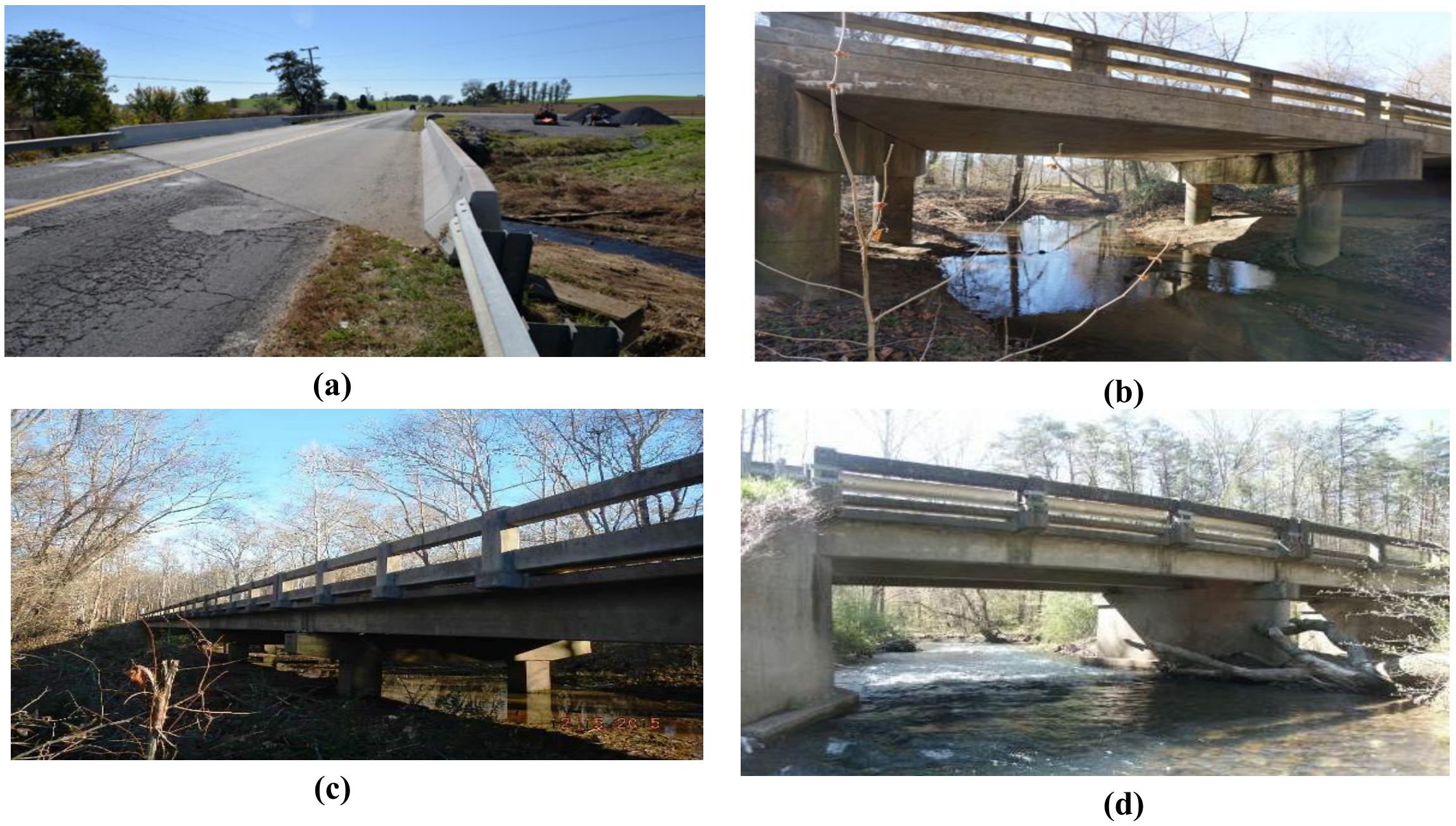

Figure 6. Target Bridges Used for Evaluating Load Rating Methodologies a) War Branch Bridge; b) Smacks Creek Bridge; c) Flat Creek Bridge; and d) Brattons Creek Bridge

## 4. Instrumentation and Field Testing

For each bridge, the instrumentation was separated into static measurement sensors and vibration sensors, but it should be noted that all of the sensors were mounted concurrently, except in cases where sensors were reconfigured for staged measurements. All instrumentation and data acquisition equipment were from Bridge Diagnostics, Inc. (BDI), and were individual sensors physically connected to four-channel nodes, which in turn interfaced wirelessly with a base station/data acquisition unit. Representative images of the instrumentation are shown in

Figure 7 and Figure 8 for illustrative purposes. Also, to compare the results obtained from updated model and measured data from live loading tests, the detail instrumentation of the two selected bridge, one slab and one T-Beam, are illustrated in Figure 9 and Figure 10.

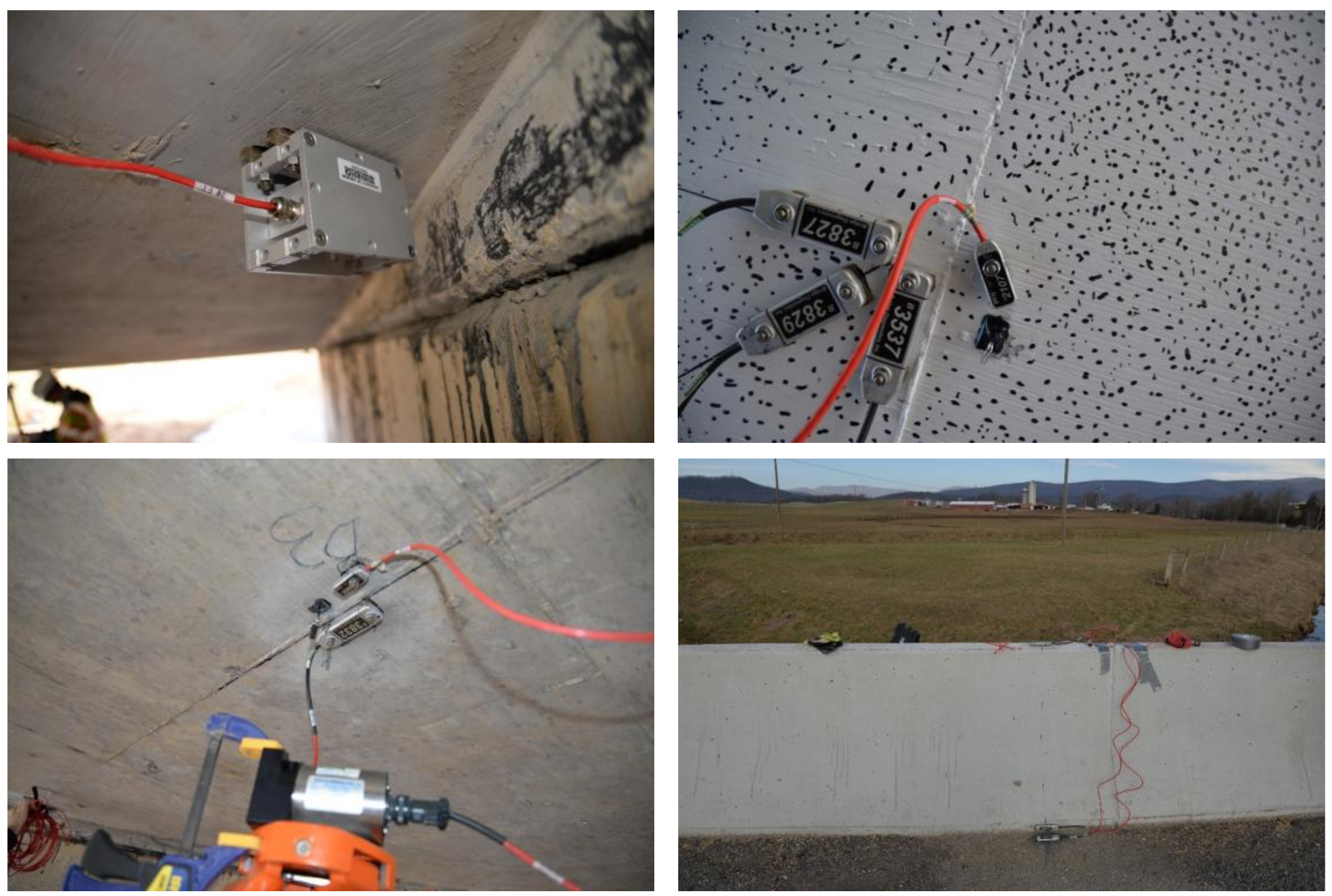

Figure 7. Representative Images of BDI Instrumentation (Clockwise from Top Left: Tiltmeter, Strain Gauge Rosette and Accelerometer, Strain Gauge and Accelerometer, Strain Gauges)

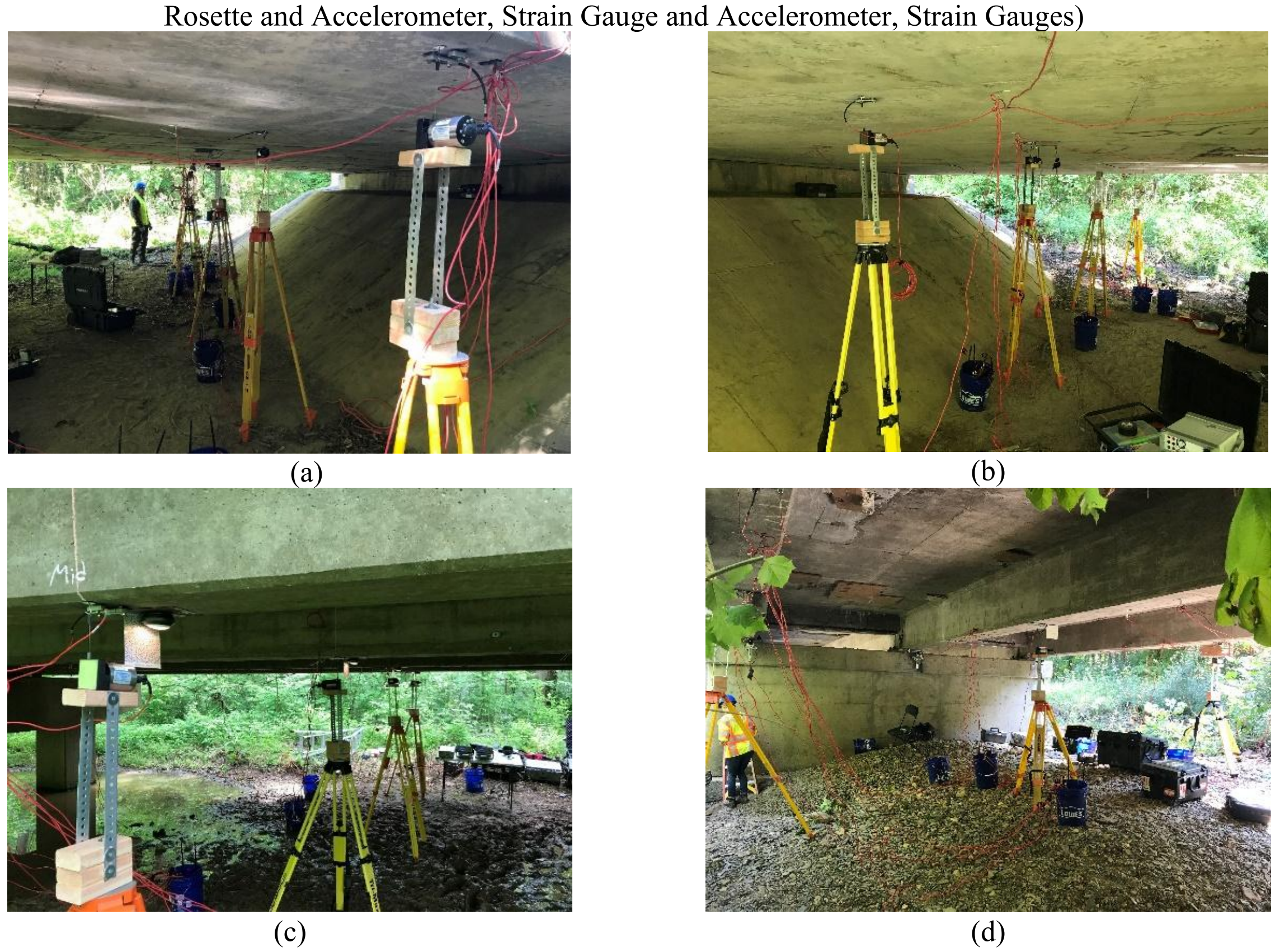

(a) (b)

(c) (d)

Figure 8 - Representative Images of BDI Instrumentation on the bridges, a) War Branch Bridge; b) Smacks Creek Bridge; c) Flat Creek Bridge; and d) Brattons Creek Bridge

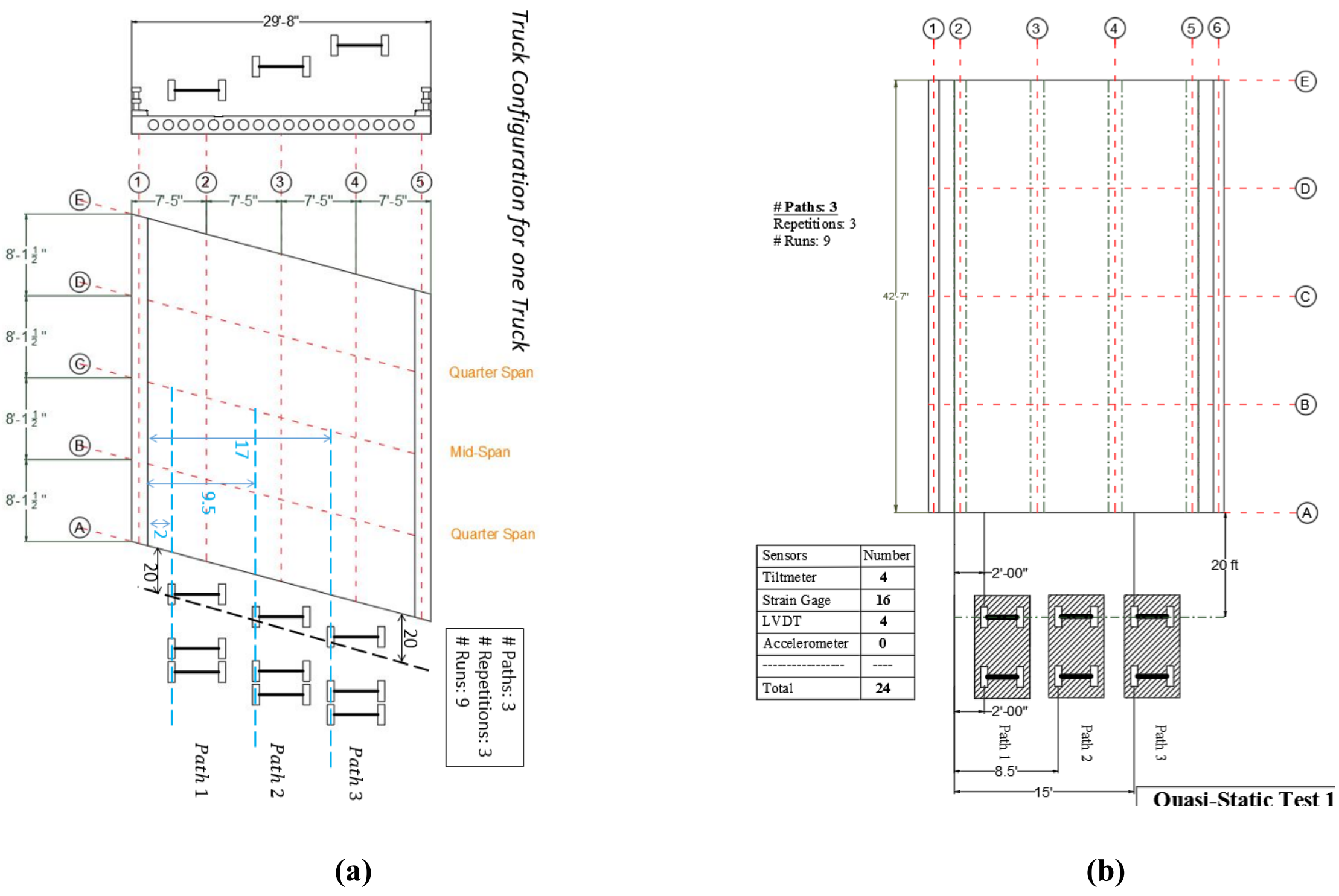

**(a)** **(b)**

Figure 9. The details of Instrumentation Configuration for live testing for different path (a) Smacks Creek, (b) Flat Creek

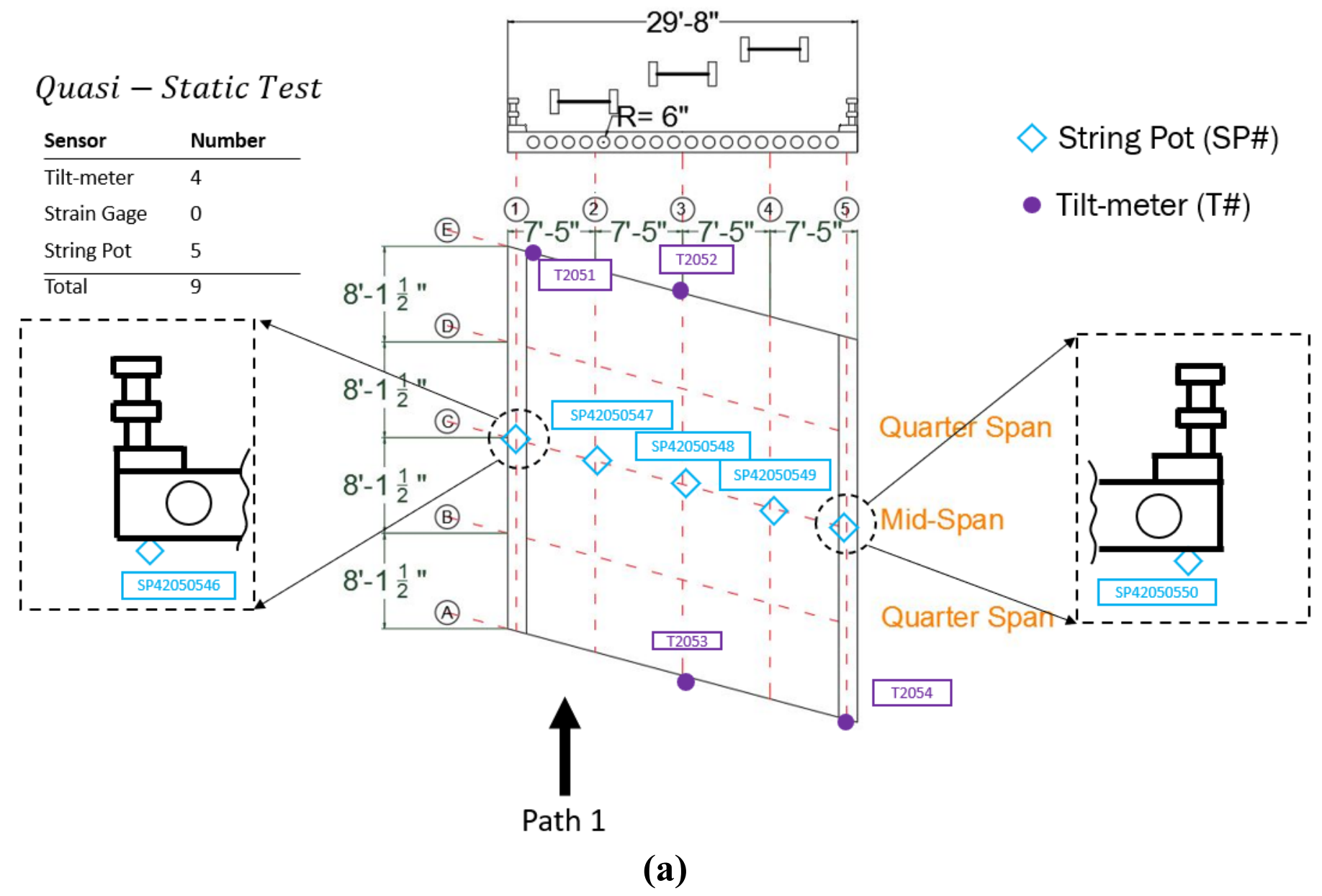

**(a)**

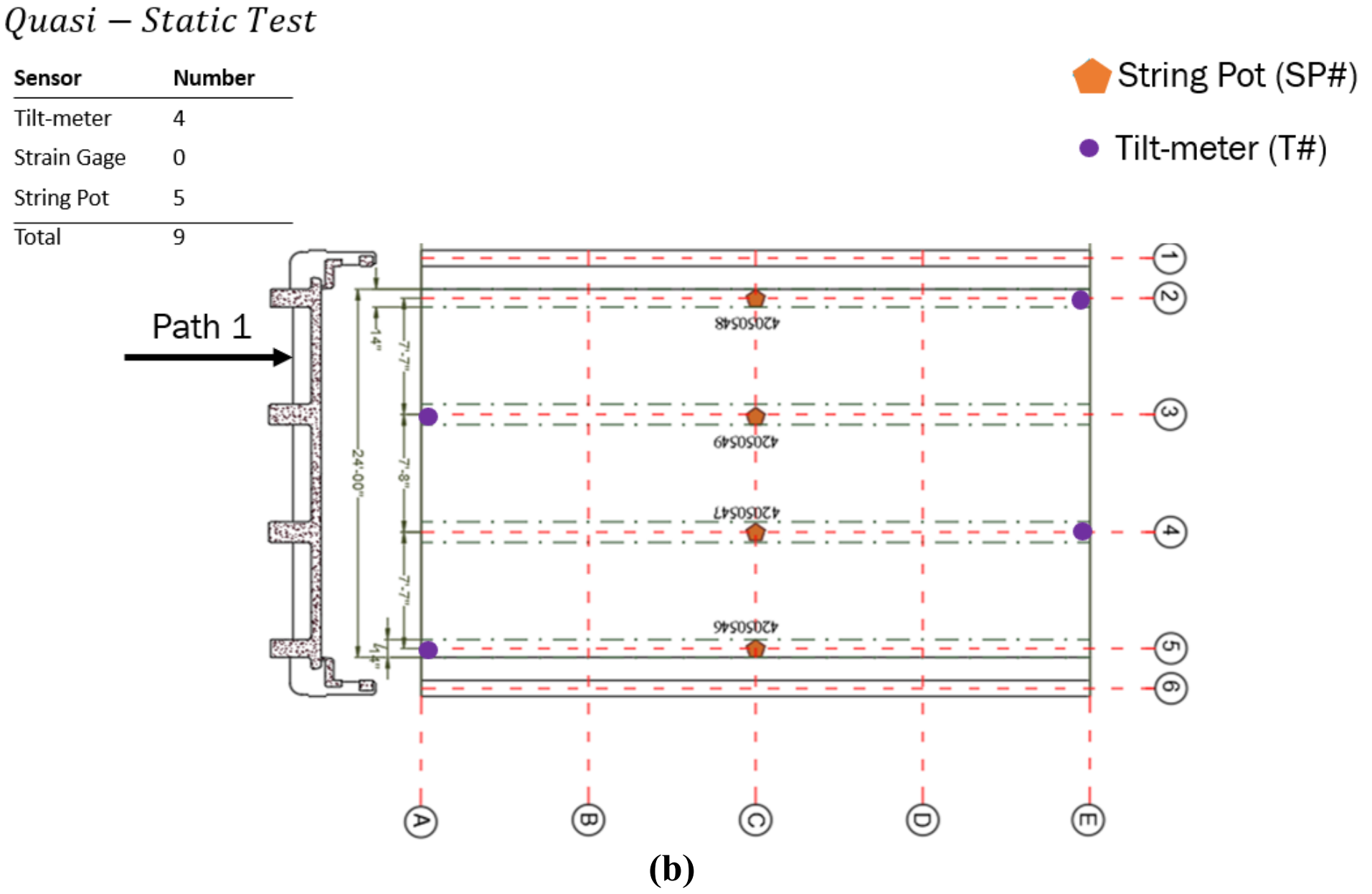

Figure 10. The details of Instrumentation Configuration for live testing (a) Smacks Creek, (b) Flat Creek

**4.1. Live Load Testing (for FEMU-S and FEMU-H scenarios)**

For the load testing experiments, each of the bridges utilized similar instrumentation and loading scenarios, with configurations tailored to the bridge geometry and configurations. In general, one of the spans in the bridges was heavily instrumented with a combination of strain gauges, string potentiometers, tiltmeters, and accelerometers with testing consisting of vibration testing and live load testing (Quasi-static and dynamic). Different vehicles were used for each of the tests, but all vehicles maintained similar configurations and comparable weights (Figure 11). The distribution of vehicle weight for each vehicle was approximately 33% on the front axle and 67% on the rear axle. Live load testing utilized a series of load configurations on the bridge under quasi-static and dynamic conditions. Live load testing experiments consisted of vehicles crossing the bridge at pre-determined transverse positions at crawl speeds (~3-5 mph), moderate speed (~25 mph), and the near the posted speed limit (~50 mph), as summarized in Table 2. Each crossing was repeated three times to ensure repeatability and reliability of the results.

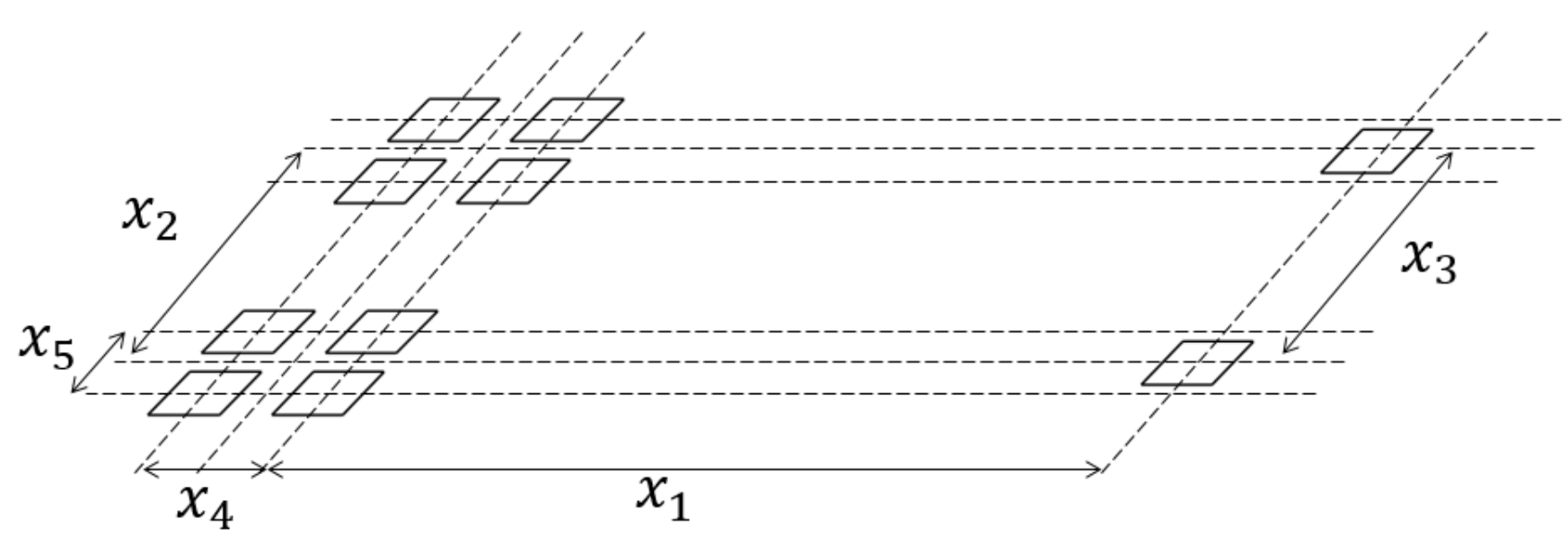

Figure 11 – Generalized Truck Configuration of Test Trucks, where $\boldsymbol{x_1}$= distance between front wheel axle and closer rear wheel axle; $\boldsymbol{x_2}$ = distance between rear wheel tires; $\boldsymbol{x_3}$ = distance between front wheel tires; $\boldsymbol{x_4}$ = distance between rear wheel axles; $\boldsymbol{x_5}$ = distance between rear tires on one side of the wheel.

Table 2 – Test Truck Details for Live Load Testing

| | **War Branch** | **Smacks Creek** | **Flat Creek** | **Brattons Creek** |
|---|---|---|---|---|
| Truck 1 | | | | |
| Gross Weight (lb) | 30,575 | 44,360 | 44,360 | 52,620 |
| $x_1$ (ft) | 12.91 | 13.5 | 13.5 | 13.5 |
| $x_2$ (ft) | 6.08 | 6 | 6 | 6 |
| $x_3$ (ft) | 6.91 | 6 | 6 | 6 |
| $x_4$ (ft) | 4 | 4 | 4 | 4 |
| $x_5$ (ft) | 1.12 | 1.12 | 1.12 | 1.12 |
| Truck 2 | | | | |
| Gross Weight (lb) | 29,600 | 45,380 | 45,380 | 52,680 |
| $x_1$ (ft) | 12.91 | 13.5 | 13.5 | 13.5 |
| $x_2$ (ft) | 6.08 | 6 | 6 | 6 |
| $x_3$ (ft) | 6.91 | 6 | 6 | 6 |
| $x_4$ (ft) | 4 | 4 | 4 | 4 |
| $x_5$ (ft) | 1.12 | 1.12 | 1.12 | 1.12 |

Note: $x_1$ - $x_5$ are defined in Fig 5

Each of the bridges tested for live load testing utilized similar instrumentation, but the configurations were established prior to testing, as shown in Figure 12. In this paper, field testing results alone are not described in detail as the focus of the investigation centers on leveraging these field test results to arrive at load ratings of the selected structures.

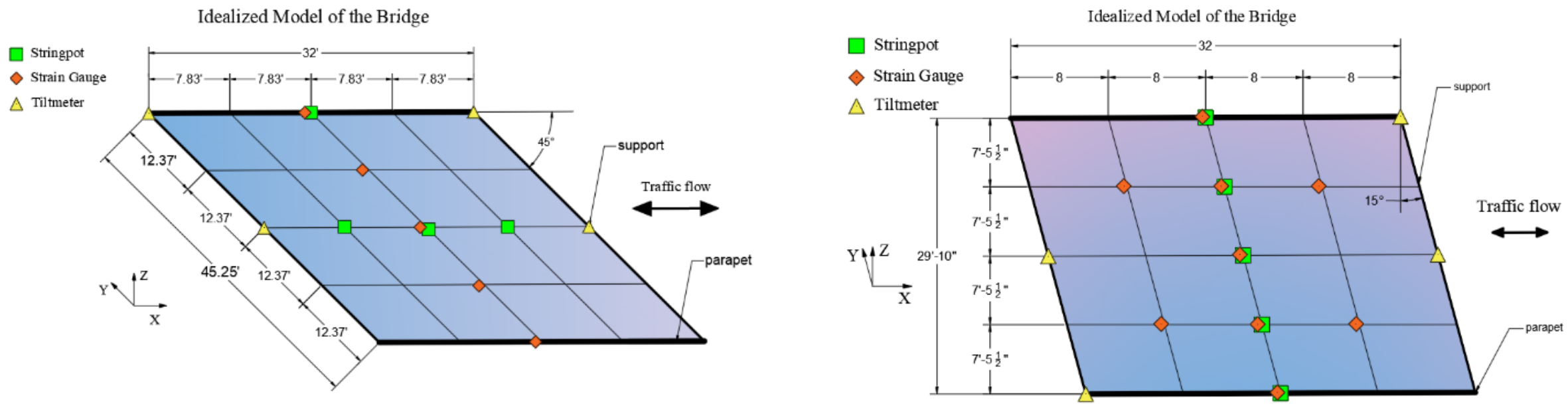

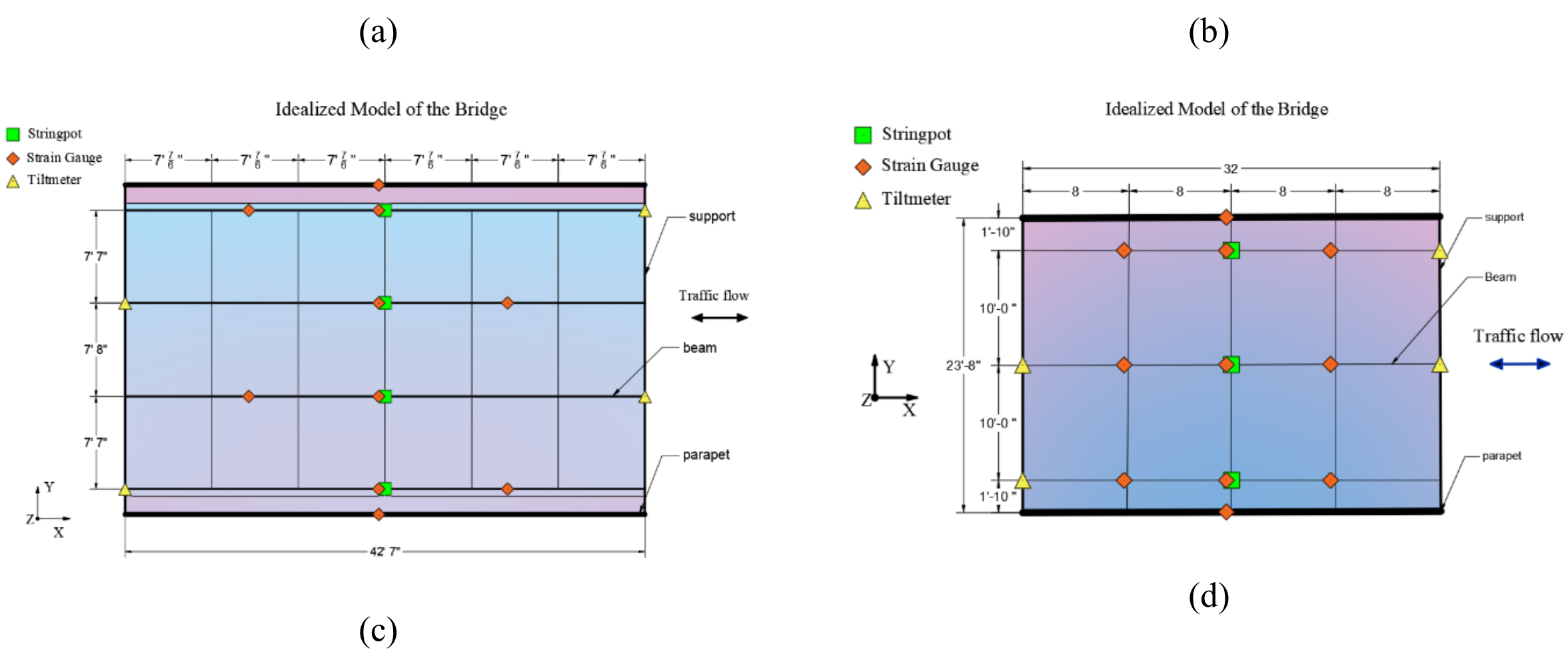

Figure 12. Instrumentation Configuration for for Live Load Testing (a) War Branch Bridge (b) Smacks Creek Branch (c) Flat Creek (d) Brattons Creek bridges

The live load testing consisted primarily of a known truck traversing the bridges at various transverse positions at controlled speeds. Representative time history plots of the raw field deflection measurements collected from string potentiometers during quasi-static and dynamic live load testing are shown in Figure 13. Results demonstrated a response that can be described as force effect and load path-guided, with the maximum response observed at a specific sensor location when the loading was driving the measured deformation.

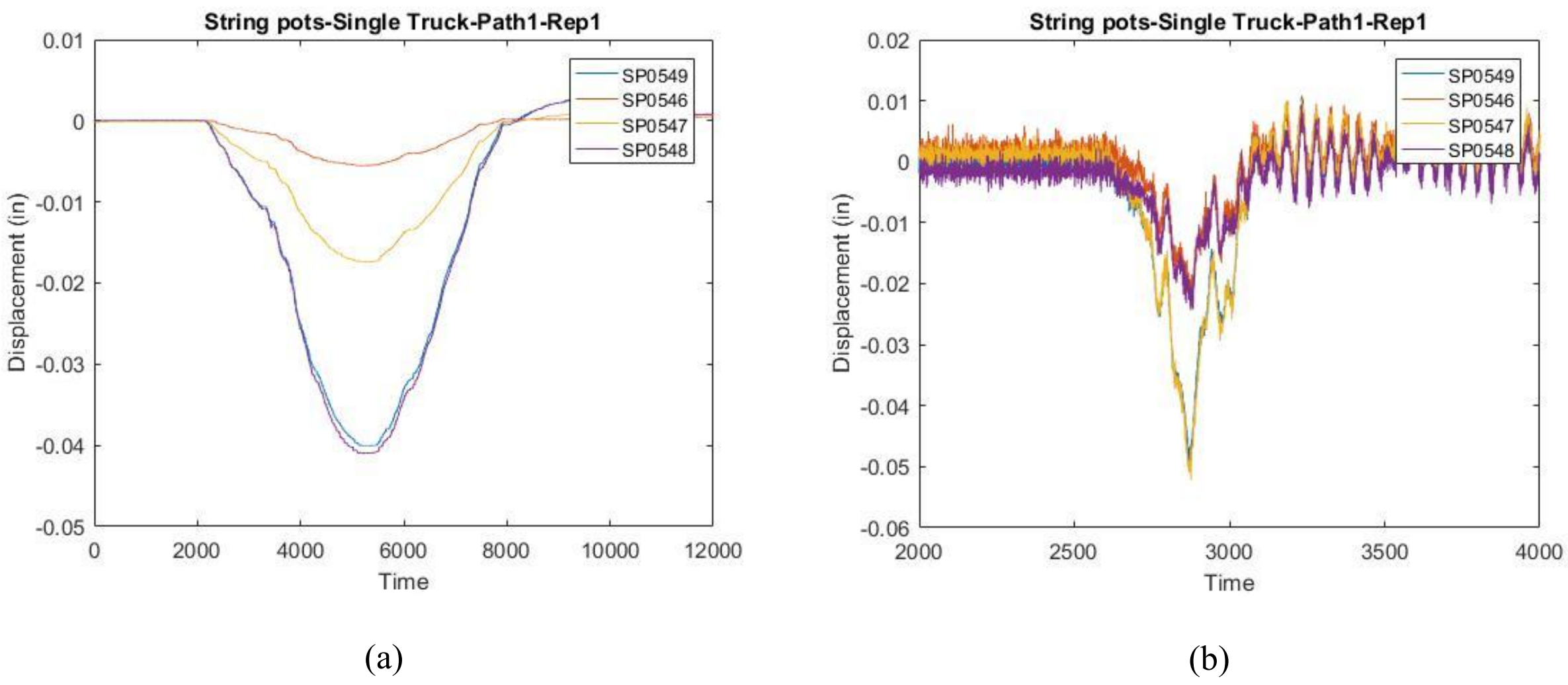

Figure 13. Representative sensor measurements (Flat Creek-T Beam Bridge 01262): (a) quasi-static live load test; (b) dynamic live load test;

### 4.2. Vibration Testing (for FEMU-D and FEMU-H scenarios):

Note that the primary goal of the vibration testing was to identify natural frequencies of the tested bridges; conducting either ambient testing or impact hammer testing provide an effective means to collect these data. Vibration testing consisted of a series of experiments with excitation provided separately by ambient loading (wind and normal traffic) and an impact hammer. Figure 14 provides a demonstration of some of the components from the vibration testing. In this project, both vibration testing methods were implemented to compare their results and provide recommendations for future implementation. For all of the vibration experiments, the tests were repeated twice; however, some of the early testing was executed in two phases in order to allow sensors to be relocated to the second half of the span once data had been collected from the first half of the span. This relocation was necessary due to the limited number of acceleration sensors available to the project team. However, for the latter tests, the spatial distribution of sensors was reduced in order to improve testing efficiency. In all tests, uniaxial accelerometers with a measuring range of ±5g were used to collect the vibration response. For the impact excitation, a large impulse sledge hammer with a force capacity range of ±5000 lbf was used. All data was collected with a sampling frequency of 500 Hz.

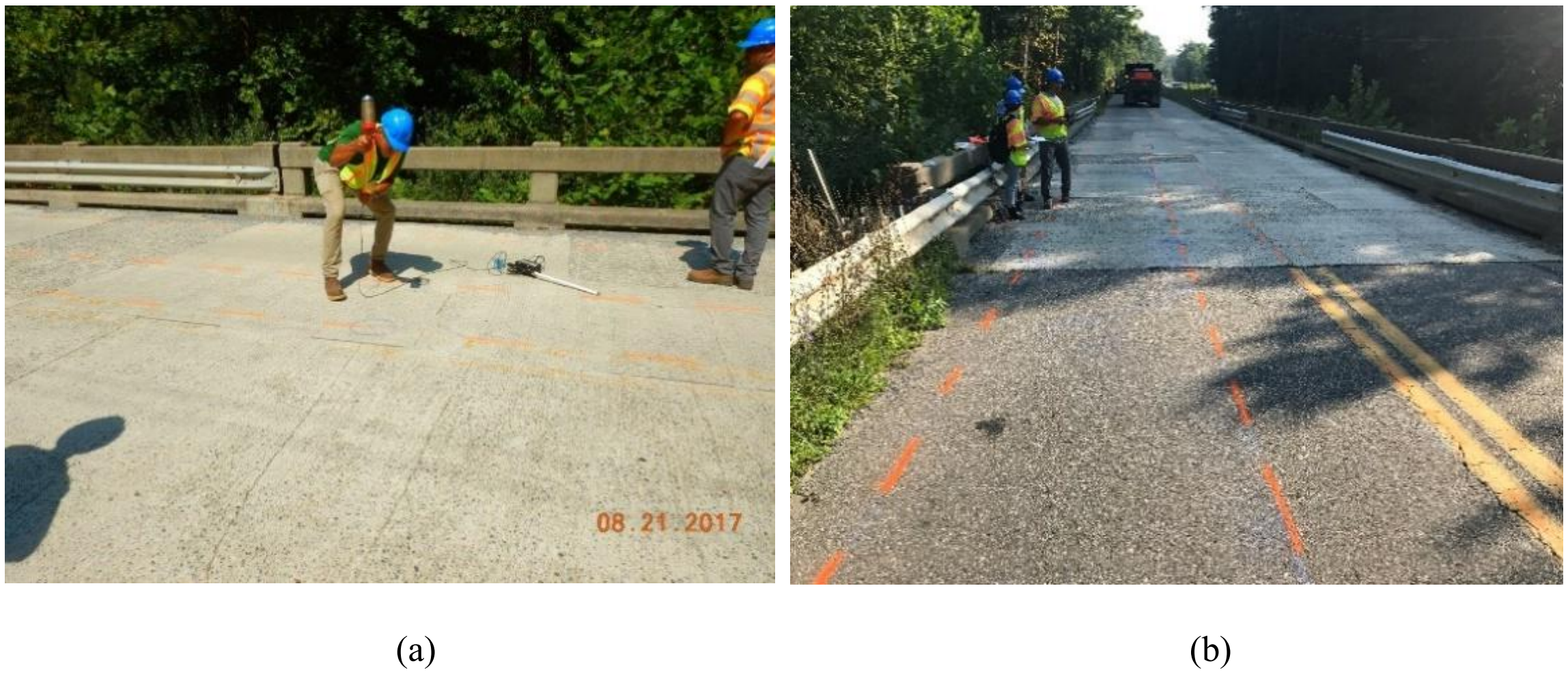

(a) (b)

Figure 14. Representation of excitation methods used during vibration testing (a) Impact Hammer Excitation; (b) Ambient Excitation

Vibration testing conducted on each selected bridge is briefly described below. This testing included both ambient excitation as well as impact hammer excitation.

War Branch Bridge (Figure 15a): The dynamic bridge assessment procedure involved the attachment of nine accelerometers underneath the selected span at 16 measurement points in two set-ups as shown in Figure 15a. In the first setup, the sensors 1 to 9 were attached to the bridge, while the sensors 10 to 18 were present on the bridge during the second setup. Note that two sensors were attached to common measurement points in each set-up and were used as reference. The response of the bridge to ambient excitations and an impact excitation were measured. Ambient vibrations were generated by the traffic, wind, and people walking across the bridge, and recorded for a total of 15 minutes. Two impact locations as shown in Figure 14 were chosen to excite the vertical/bending modes of the bridge using the modal impact hammer. These chosen points were excited by the impact hammer for five repetitions and a duration of 15 seconds.

Smacks Creek Bridge (Figure 15b): The dynamic bridge assessment procedure involved the attachment of accelerometers underneath one span of the bridge at 9 measurement points in one set-up, as shown in Figure 15b. The response of the bridge to ambient excitations or an impact excitation generated by a modal impact hammer was measured. Ambient vibrations were generated by the passing traffic, wind and walking people and recorded for a total of 15 minutes. Three impact locations were chosen and excited by the impact hammer for five repetitions and for a duration of 15 seconds.

Flat Creek Bridge (Figure 15c): The dynamic bridge assessment procedure involved the attachment of accelerometers underneath one span of the bridge for a total of 19 measurement points in two set-ups, as shown in Figure 15c. In the first set up, the sensors 1 to 10 were attached to the bridge, while the sensors 11 to 20 were present on the bridge during the second set-up. One sensor was present in a common measurement point in both set-ups and was used as reference point. The connection of the accelerometers to the girders of the deck was performed by means of metallic plates bonded to the surface of the concrete. Note that two common sensors were used as reference in each set-up. Ambient vibrations were generated by the passing traffic, wind and walking people and recorded for a total of 15 minutes.

Brattons Creek Bridge (Figure 15d): The dynamic bridge assessment procedure involved the attachment of accelerometers underneath the span of the bridge at 9 measurement points in a single set-up, as shown in Figure 15d. The connection of the accelerometers to the girders of the deck was performed by means of metallic plates bonded to the surface of the concrete. Note that two common sensors were used as reference in each set-up. Ambient vibrations were generated by the passing traffic, wind and walking people and recorded for a total of 15 minutes.

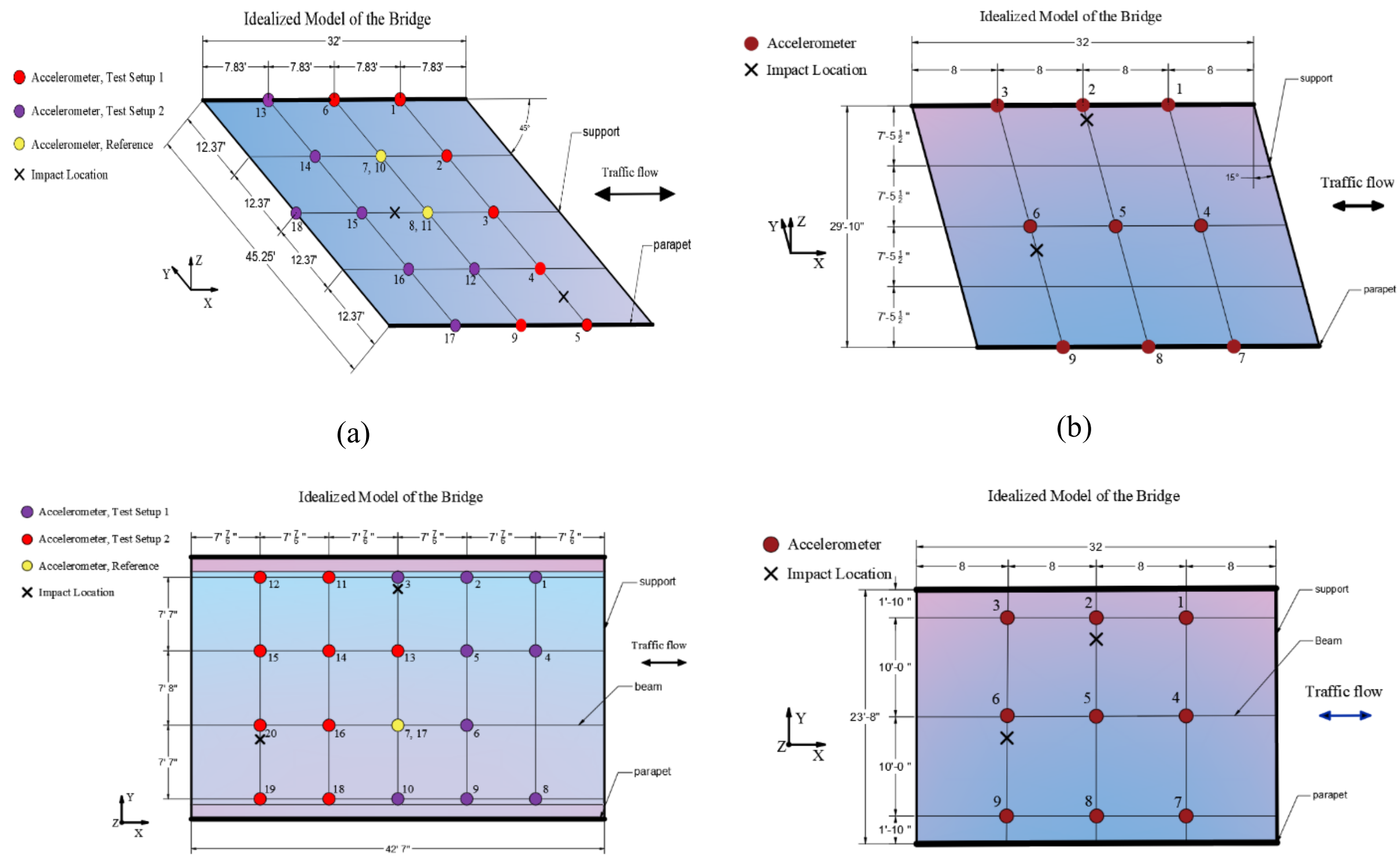

Figure 15. Instrumentation Configuration for Vibration Testing (a) War Branch Bridge (b) Smacks Creek, (c) Flat Creek (d) Brattons Creek Bridge

Figure 16**Error! Reference source not found.** illustrates representative time series plots of acceleration data collected during the ambient vibration and impact hammer experiments. It can be seen that peak acceleration responses were observed as a result of either large magnitude ambient loadings (vehicle passing) or impact from the hammer strike. Table 3 and Table 4 provide summaries of the modal properties identified for each of the structures using the previously described EFDD approach for both the ambient and impact excitation methods.

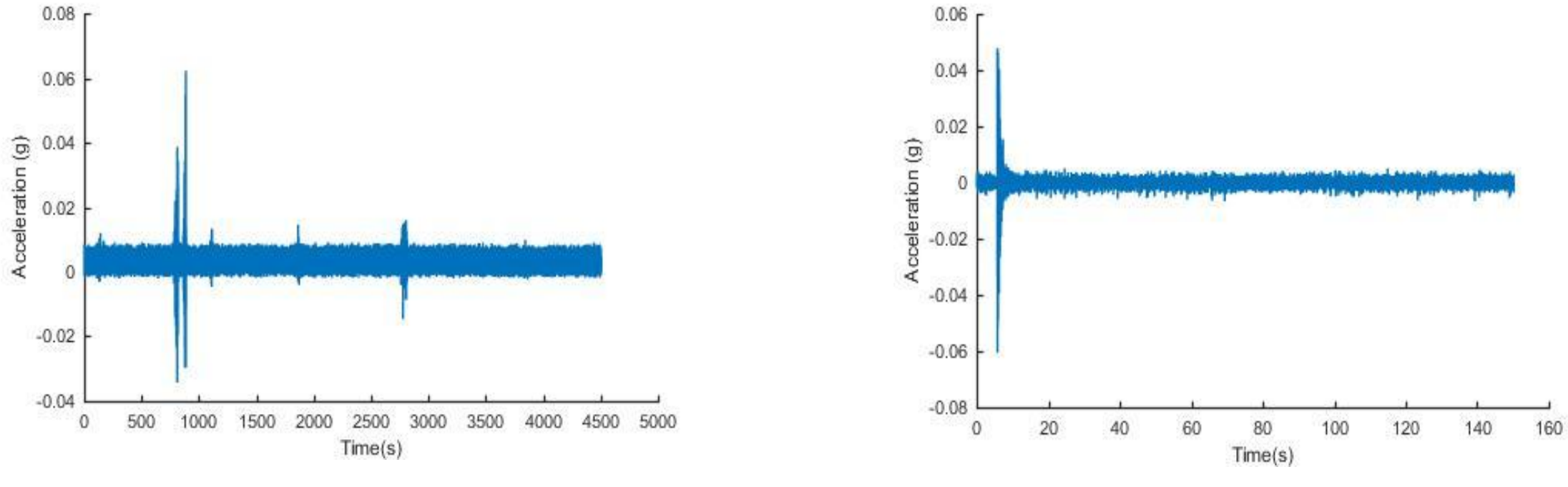

(a) (b)

Figure 16. Representative Sensor Measurements (Smacks Creek-T Beam Bridge): (a) Ambient Vibration; and (b) Impact Hammer Vibration

Table 3 – Summary of Identified Modal Properties from Ambient Vibration Excitation

| | Modal Frequency (Hz) | | | | Damping Ratio % | | | |
|---|---|---|---|---|---|---|---|---|
| **Mode** | **War Branch** | **Smacks Creek** | **Flat Creek** | **Brattons Creek** | **War Branch** | **Smacks Creek** | **Flat Creek** | **Brattons Creek** |
| 1 | 25.78 | 14.11 | 10.74 | 14.24 | 2.92 | 5.029 | 6.061 | 1.97 |
| 2 | 32.22 | 20.007 | 13.77 | 17.53 | 1.64 | 3.569 | 0 | 0.76 |
| 3 | 38.41 | 32.70 | 18.54 | 25.06 | 2.69 | 4.44 | 2.36 | 0.54 |

Table 4 - Summary of Identified Modal Properties from Impact Hammer Excitation

| | Modal Frequency (Hz) | | | | Damping Ratio % | | | |
|---|---|---|---|---|---|---|---|---|
| **Mode** | **War Branch** | **Smacks Creek** | **Flat Creek** | **Brattons Creek** | **War Branch** | **Smacks Creek** | **Flat Creek** | **Brattons Creek** |
| 1 | 26.35 | 14.08 | 11.09 | 14.15 | 4.17 | 2.76 | 7.85 | 1.94 |
| 2 | 32.72 | 20.03 | 13.94 | 17.32 | 4.16 | 2.32 | 1.76 | 1.52 |
| 3 | 39.06 | 32.63 | 18.40 | 22.99 | 2.71 | 2.64 | 3.67 | 1.38 |

### 4.3. Modal Parameter Identification (for FEMU-D and FEMU-H scenarios):

Operational modal analysis enables the derivation of the modal parameters from the dynamic response of a structure under operational loads. A number of methods have been developed for output-only system identification of structures. In this study, the Enhanced Frequency Domain Decomposition (EFDD) algorithm was used for identifying the modal properties of the bridge from the collected acceleration response. In the simple FDD method, unknown inputs and acquired outputs are related through their power spectral densities and frequency response functions [27]. By processing the outputs from the experimental data, the power spectral density matrix is estimated. The output power spectral density at discrete frequencies is then decomposed by taking the singular value decomposition of the matrix. The corresponding singular value is the power spectral density function of the single degree of freedom system. This power spectral density function is identified by isolating the peak and comparing the mode shape estimate with the singular vectors obtained for frequency lines around the peak. Enhanced frequency domain decomposition method is an extension to frequency domain decomposition (FDD) method. As FDD method is based on using a single frequency line from the Fast Fourier Transform (FFT), the accuracy of the estimated natural frequency depends on the FFT resolution and no modal damping is calculated. However, EFDD method [28] gives an improved estimation of both the natural frequencies and the mode shapes while also including the damping ratios. The EFDD technique allows us to extract the resonance frequency and the damping of a particular mode by computing the auto and cross-correlation functions. The Single Degree of Freedom (SDOF) power spectral density function, identified around a peak of resonance, is taken back to the time domain using the Inverse Discrete Fourier Transform. The resonance frequency is obtained by determining the zero crossing times and the damping by the logarithmic decrement of the corresponding SDOF normalized auto correlation function. For all of the tested bridges, the modal parameters were identified using the EFDD applied to both ambient and impact vibration data. In this study, the implementation of the EFDD method available in the ARTeMIS Modal Pro software was used for modal parameter identification [29].

## 5. Load Rating Results from Different Methodologies

*AASHTO LRFR Load Rating*. As a baseline to which the results of all the other methods could be compared, the conventional analytical load rating method described in the AASHTO MBE was performed [1]. In particular, the LRFR method, which is the most current method and corresponds to the AASHTO LRFD Bridge Design Specifications [2], was selected as the baseline in this investigation. All the structural and mechanical properties of the bridges according to the plan of the bridges are shown in Table 5, where: $\rho$ = Density of concrete, $E_c$ = Elasticity modulus of concrete, $f_y$ = Yield strength of reinforcing steel, $f_c^{'}$ = Compressive strength of concrete, $M_n$ = Nominal bending capacity, $A_s$ = Area of reinforcing steel. For the slab bridges, $A_s$ is calculated for a foot-wide section and for T -beam bridges, $A_s$ is calculated for one beam, while Table 5 shows the values for a single bar within the cross-section. For instance, in the War Branch Bridge cross-section within every one foot there are two rebar with area of steel 1.27 in$^2$; therefore, the entire area of steel within this one-foot strip would be 2.54 in$^2$. Table 5 presents a summary of the results for all of the bridges evaluated using the AASHTO LRFR method.

Table 5 - Properties of the Bridges Based on As-built Plans and the results

| Properties | War Branch | Smacks Creek | Flat Creek | Brattons Creek |
|---|---|---|---|---|
| $\rho\ (pcf)$ | 150 | 150 | 150 | 150 |
| $E_c\ (ksi)$ | 3,322 | 3,322 | 3,322 | 3,322 |
| $A_s (in^2)$ | 1.27 | 1.27 | 1.27 | 1.27 |
| $f_y (ksi)$ | 40 | 40 | 40 | 40 |
| $f_c^{'} (ksi)$ | 3.00 | 3.00 | 3.00 | 3.00 |
| AASHTO LRFR ($RF_c$) | 1.37 | 1.07 | 0.88 | 0.71 |
| Through Diagnostic Load Testing ($RF_c$) | 1.38 | 1.06 | 0.85 | 0.71 |

*AASHTO Load Rating Through Diagnostic Load Testing*. Based on the experimental data collected during the field testing, the diagnostic testing allowed for the original load rating factors, $RF_c$, to be adjusted to a higher values ($RF_T$). Table 5 provides a summary of the ratings derived from the AASHTO diagnostic testing method. The results demonstrated the potential improvement in load rating that can be derived from a better understanding on the as-built response and performance versus design approximation.

## 6. Finite Element Model Updating based Load Rating

Results for the estimated parameters, calculated capacity, and derived load rating factors are summarized in Table 6 to Table 9 for each of the bridges evaluated. For the FEMU-S and FEMU-H methods, the results are presented for three different loading configurations (i.e. Paths 1 through 3 shown in Figure 9) to illustrate the robustness of the approach. The comparison of the results from the updated model and measured data for the selected bridges shows that the proposed FE updating approach was able to fine-tune the initial model effectively shown in Figure 17 and Figure 18.

Table 6 –Estimated Parameters and Load Ratings based on FEMU Approaches (War Branch)

| **Parameter / Result** | **FEMU-S** | | | **FEMU-D** | **FEMU-H** | | |
|---|---|---|---|---|---|---|---|
| | Path 1 | Path 2 | Path 3 | N/A | Path 1 | Path 2 | Path 3 |
| $E_c\ (ksi)$ | 4,502 | 4,510 | 4,322 | 5,003 | 4,288 | 3,744 | 4,115 |
| $A_s(in^2)$ | 1.30 | 1.30 | 1.30 | 1.40 | 1.20 | 1.20 | 1.20 |
| $f_c'(ksi)$ | 4.68 | 4.87 | 4.68 | 5.79 | 4.25 | 3.24 | 3.91 |
| $M_n(kip-ft)$ | 150.90 | 150.94 | 150.18 | 163.81 | 139.14 | 136.38 | 138.38 |
| $RF_{moment}\ (Inventory)$ | 1.52 | 1.99 | 1.51 | 1.69 | 1.36 | 1.33 | 1.36 |
| $RF_{moment}\ (Operating)$ | 2.27 | 2.26 | 2.25 | 2.51 | 2.04 | 1.98 | 2.02 |
| $RF_{VDOT}\ (Inventory)$ | 1.39 | | | | | | |

Table 7 – Estimated Parameters and Load Ratings based on FEMU Approaches (Smacks Creek)

| **Parameter / Result** | **FEMU-S** | | | **FEMU-D** | **FEMU-H** | | |
|---|---|---|---|---|---|---|---|
| | Path 1 | Path 2 | Path 3 | N/A | Path 1 | Path 2 | Path 3 |
| $E_c\ (ksi)$ | 3,084 | 3,200 | 3,115 | 4,335 | 3,988 | 3,899 | 4,122 |
| $A_s(in^2)$ | 1.21 | 1.21 | 0.94 | 0.96 | 1.10 | 1.11 | 1.22 |
| $f_c'(ksi)$ | 2.21 | 2.37 | 2.24 | 4.34 | 3.67 | 3.51 | 3.93 |
| $M_n(kip-ft)$ | 131.83 | 133.07 | 105.61 | 112.85 | 127.07 | 130.52 | 140.56 |
| $RF_{moment}\ (Inventory)$ | 0.98 | 0.99 | 0.70 | 0.78 | 0.92 | 0.92 | 1.06 |
| $RF_{moment}\ (Operating)$ | 1.46 | 1.48 | 1.07 | 1.18 | 1.39 | 1.39 | 1.58 |
| $RF_{VDOT}\ (Inventory)$ | 1.05 | | | | | | |

Table 8 – Estimated Parameters and Load Ratings based on FEMU Approaches (Flat Creek)

| Parameter / Result | FEMU-S | | | FEMU-D | FEMU-H | | |
|---|---|---|---|---|---|---|---|
| | Path 1 | Path 2 | Path 3 | N/A | Path 1 | Path 2 | Path 3 |
| $E_c\ (ksi)$ | 3,301 | 3,341 | 3,479 | 4,100 | 3,114 | 3,254 | 3,098 |
| $A_s(in^2)$ | 0.80 | 1.00 | 0.91 | 1.00 | 1.10 | 1.10 | 1.20 |
| $f'_c(ksi)$ | 2.52 | 2.58 | 2.79 | 3.88 | 2.24 | 2.45 | 2.22 |
| $M_n(kip-ft)$ | 798.26 | 992.95 | 897.79 | 1,062.03 | 1,084.32 | 1,087.49 | 1,178.69 |
| $RF_{moment}\ (Inventory)$ | 0.43 | 0.63 | 0.53 | 0.59 | 0.72 | 0.73 | 0.82 |
| $RF_{moment}\ (Operating)$ | 0.67 | 0.96 | 0.82 | 0.77 | 1.09 | 1.10 | 1.24 |
| $RF_{VDOT}\ (Inventory)$ | 0.84 | | | | | | |

Table 9 – Estimated Parameters and Load Ratings based on FEMU Approaches (Brattons Creek)

| Parameter / Result | FEMU-S | | | FEMU-D | FEMU-H | | |
|---|---|---|---|---|---|---|---|
| | Path 1 | Path 2 | Path 3 | N/A | Path 1 | Path 2 | Path 3 |
| $E_c\ (ksi)$ | 3,711 | 4,113 | 4098 | 3,666 | 3,601 | 3,887 | 3,741 |
| $A_s(in^2)$ | 1.68 | 1.63 | 1.51 | 1.40 | 1.23 | 1.20 | 1.25 |
| $f'_c(ksi)$ | 3.18 | 3.91 | 2.79 | 3.10 | 2.99 | 3.49 | 3.23 |
| $M_n(kip-ft)$ | 903.87 | 882.36 | 820.54 | 754.95 | 666.50 | 650.88 | 698.30 |
| $RF_{moment}\ (Inventory)$ | 1.03 | 1.00 | 0.91 | 0.81 | 0.68 | 0.66 | 0.59 |
| $RF_{moment}\ (Operating)$ | 1.63 | 1.58 | 1.44 | 1.30 | 1.10 | 1.07 | 0.76 |
| $RF_{VDOT}\ (Inventory)$ | 0.77 | | | | | | |

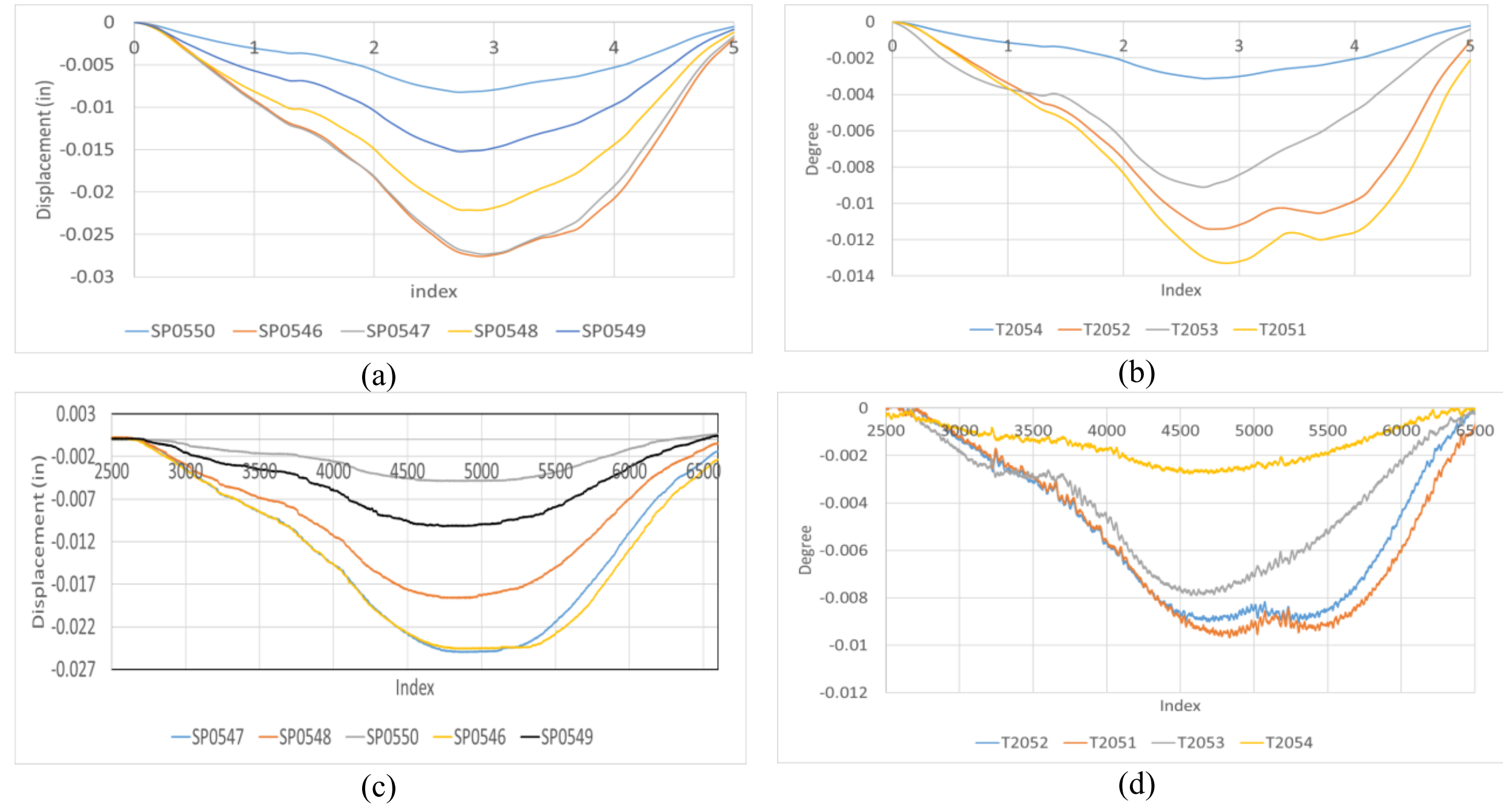

Figure 17. Comparison of the results from updated model using FEMU-H and measured data for the Smacks Creek Slab Bridge, Path 1, (a) updated displacements, (b) updated rotations, (c) measured displacements, (d) measured rotations

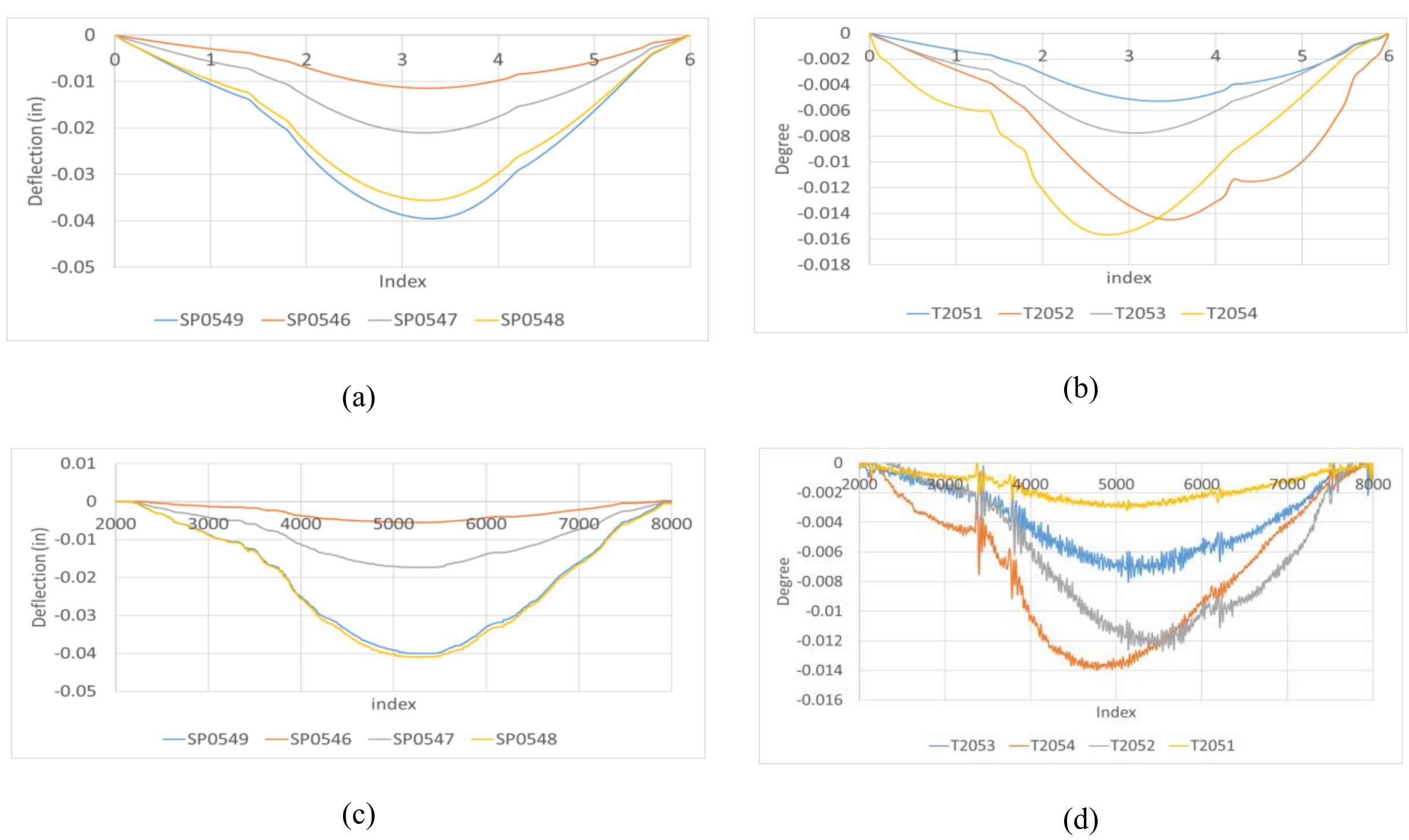

Figure 18. Comparison of the results from updated model using FEMU-H and measured data for the Flat Creek T-Beam bridge, Path 1, (a) updated displacements, (b) updated rotations, (c) measured displacements, (d) measured rotations.

## 7. Comparison and Discussion of Load Rating factors Using Different Methods

Table 10 presents a summary of the various load rating methods explored in this study. The comparison uses the Inventory load rating factor for comparison, but the Operating load rating factor would give a similar result. For the FEMU-S and FEMU-H methods, the table presents the results from each loading configurations (i.e. truck paths) as well as the average result obtained from these tests.

Table 10 –Inventory Load Rating Results from Different Analyses

| Rating Method | | War Branch | Smacks Creek | Flat Creek | Brattons Creek |
|---|---|---|---|---|---|
| AASHTO LRFR ($RF_c$) | | 1.38 | 1.06 | 0.85 | 0.71 |
| AASHTO Diagnostic ($RF_T$) | | 1.38 | 1.06 | 0.85 | 1.02 |
| FEMU-S | Path 1 | 1.52 | 0.98 | 0.43 | 1.03 |
| | Path 2 | 1.99 | 0.99 | 0.63 | 1.00 |
| | Path 3 | 1.51 | 0.70 | 0.53 | 0.91 |
| | Avg.* | 1.67 | 0.89 | 0.53 | 0.98 |
| FEMU-D | | 1.69 | 0.78 | 0.59 | 0.81 |
| FEMU-H | Path 1 | 1.36 | 0.92 | 0.72 | 0.68 |
| | Path 2 | 1.33 | 0.92 | 0.73 | 0.66 |
| | Path 3 | 1.36 | 1.06 | 0.82 | 0.59 |
| | Avg.* | 1.35 | 0.97 | 0.76 | 0.64 |

* Average of three truck paths

In evaluating the results, the proposed methodologies are able to provide reasonable estimates of the rating factors relative to those derived using the baseline AASHTO LRFR method. In this case, "reasonable" refers to rational estimates on the same order of magnitude; in some cases, these estimates are very close to the AASHTO LRFR estimates. Ideally the proposed load rating method would be able to yield conservative approximations of load rating ($RF_{method} < RF_{AASHTO}$) that may not be achieved with subjective rating practices. However, for the tested structures, this outcome was not observed consistently for each load rating method evaluated. Nevertheless, the proposed method provides a significantly more objective means of calculating a rating factor for bridges with missing nformation as opposed to subjective and judgement-driven alternative used in current practice.

As shown in Table 11, the results obtained from FEMU-S method, which relies on static measurement for model updating, produced the largest rating differences compared to the RF obtained from the AASHTO LRFR method. For two of the tested bridges, the method overestimated the rating factor while for other two bridges it underestimated the rating factor. In addition, for a given bridge, the results obtained from this method showed a degree of loading configuration (truck path) dependence. This can be associated with localized effects such as uneven and non-uniform material distribution, boundary conditions, and rebar configurations. As a result, it is recommended to repeat the quasi-static live-load test with a few load paths and use the average load ratings to produce more robust load rating estimates.

Table 11 – Comparison of Percent Difference* Between Method Load Ratings and AASHTO Load Ratings

| **Rating Method** | | **War Branch** | **Smacks Creek** | **Flat Creek** | **Brattons Creek** |
|---|---|---|---|---|---|
| FEMU-S | Path 1 | 10% | -8% | -49% | 45% |
| | Path 2 | 44% | -7% | -26% | 41% |
| | Path 3 | 9% | -34% | -38% | 28% |
| | Avg.** | 21% | -16% | -38% | 38% |
| FEMU-D | | 22% | -26% | -31% | 14e % |
| FEMU-H | Path 1 | -1% | -13% | -15% | -4% |
| | Path 2 | -4% | -13% | -14% | -7% |
| | Path 3 | -1% | 0% | -4% | -17% |
| | Avg.** | **-2**% | -9% | **-11**% | -9% |

* Percent difference defined as $(RF_{method} - RF_C)/RF_C * 100\%$

** Average of three runs

The results derived from the FEMU-H method, which includes both static and dynamic measurements in model updating, produced the most reasonable estimates of rating factors with a percent difference ranging from 0% to -17%, with negative differences indicating lower estimates of RF than the AASHTO standard. Using multiple load paths and the resulting average load ratings, the discrepancy of the estimates is reduced to 2-11% for the four bridges. Based on the results, the FEMU-H load rating method tends to consistently estimate load ratings less than those derived from the AASHTO LRFR method, suggesting an overall degree of conservativeness; however, it should be emphasized that only four bridges were evaluated and further testing and analysis is required to determine the generalization of these results to all bridges. In addition, a significantly reduced degree of differences in the load rating estimates among different loading configurations is observed when using the hybrid scenario compared with the static-only counterpart, indicating the improved robustness of the method in terms of testing configuration. The results obtained from the FEMU-D method, which used only dynamic measurements for the model updating, produced better results than FEMU-S method but not as good and consistent as those obtained from the FEMU-H method.

Note that the FEMU-H method requires the identification of natural frequencies of the bridge structure. In this study, vibration testing with both ambient excitations, which does not require any traffic control, and impact hammer excitation, which requires at least partial traffic closure, were considered. As shown in Tables 4 and 5, the natural frequencies obtained from ambient vibration testing and impact hammer testing were very close to each other for all four tested bridges. This indicates that ambient vibration testing where the bridge is excited by passing traffic can be reliably used as the preferred vibrating testing method as it has minimal effects on the operational condition of the bridge.

### 8.1. Sensitivity Analysis of Finite Element Model Updating-based Methods

In traditional bridge testing, there are a variety of sensors that can be used to describe the static/dynamic behaviors of in-service structures. The general assumption in load and modal testing of bridges is that the more measured data collected from the structure, the better the numerical model will be in tune with the structure. However, installing more sensors can be difficult, especially in complex structures, but also uneconomical in terms of time and money. Selecting an appropriate sensor configuration becomes very important to achieving a satisfactory model updating result since the sensors are directly related to the change of the static/dynamic properties predicted by FE models.

Therefore, selection of an optimal number/type of sensors to sufficiently characterize the static/dynamic behavior of a bridge remains a topic of study.

In this work, the original sensor selection and placement criteria was based on previous experiences in live load and modal testing on other bridges. The primary goal in the selection process was characterization of expected behaviors from the tested structures, such as load sharing, composite action, boundary restraint, operational mode shapes, and damping effects. This resulted in a relatively dense sensor placement during testing; however, the expectation was that these measurements could be reduced for the purposes of implementation. To evaluate the impacts of the reduction in sensor usage, two scenarios were evaluated to assess the performance of the FEMU-H scenario in estimating the unknown parameters of interest. The reductions focused on selecting sensor measurement combinations that were likely to be used in traditional live load or modal testing, rather than the more extensive distributed sensing approach used in the method development. To achieve this outcome, the following optimization processes were evaluated:

- *Scenario 1: Using midspan deflection and accelerometer sensors* . This type of data is typically collected during a traditional live load test. It is used to describe global load sharing amongst components of the bridge along with the global vibration characteristics.
- *Scenario 2: Using midspan longitudinal strain and accelerometer sensors* . This type of data can be collected during a traditional live load test to describe both load sharing behavior and localized member deformation along with the global vibration characteristics.

For the scenarios described, the model updating processes was carried out and the unknown parameters were estimated. A summary of the results is provided in Table 12 along with the FEMU-H results using all sensors previously described (bold values indicate largest errors). The results do not clearly indicate that any of the sensing configurations are more appropriate than the others, but do illustrate that comparable results can be achieved using a more limited sensor suite. The results suggest that the updating process is not constrained by the sensor configuration. However, a general principle within a St-ID framework is the assurance that the measurements used in a model updating strategy are inclusive of response characteristics that are activated during testing (e.g. deflection measurements at midspan during a live load test, vertical accelerations for flexural impact test, or support rotation measurements for a midspan loading).

Table 12 - Summary of Sensitivity Analysis Results for War Branch Bridge

| **Scenario** | $E_c$ | $E_c/E_{c0}$ | $A_s$ | $A_s/A_{s0}$ | $E_{c\text{-error}}$ (%) | $A_{s\text{-error}}$ (%) |
|---|---|---|---|---|---|---|
| *Path 1* | | | | | | |
| 1 | 3,988 | 1.10 | 1.0 | 0.79 | 10 | **21** |
| 2 | 3,111 | 0.86 | 1.4 | 1.10 | **14** | 10 |
| FEMU-H | 4,288 | 1.18 | 1.2 | 0.94 | 29 | 6 |
| *Path 2* | | | | | | |
| 1 | 3668 | 1.01 | 1.2 | 0.94 | 1 | 6 |
| 2 | 3899 | 1.08 | 1.1 | 0.86 | 8 | 14 |
| FEMU-H | 3,744 | 1.04 | 1.2 | 0.94 | 4 | 6 |
| *Path 3* | | | | | | |
| 1 | 3974 | 1.10 | 1.2 | 0.94 | 10 | 6 |
| 2 | 3110 | 0.86 | 1.4 | 1.10 | 14 | 10 |
| FEMU-H | 4,115 | 1.14 | 1.2 | 0.94 | 14 | 6 |

## 8. Conclusions

This paper focused on developing a rational methodology for determining the load rating of bridges with insufficient details or unknown plans. The proposed methodology involves a FE model updating scheme that uses sensing data from live-load and/or vibration tests to iteratively hone in on internal structural properties that are needed for a load rating analysis. Three scenarios were defined where the optimization used either static-only or vibration-only sensing data or a combination of the two. It was demonstrated that the updating process is enhanced with the inclusion of measurements that capture both the local (static) and global (vibration) characteristics of the bridge under consideration. The outcomes of the study highlighted a general approach that is suitable for estimating load ratings in the absence of sufficient details. Based on the results obtained in this study, the following conclusions can be made:

- The estimated load ratings for four test bridges showed the feasibility of the proposed FEMU-based method. The results derived from the FEMU-H method, which includes both static and dynamic measurements in model updating, produced the most reasonable estimates of rating factors with a percent difference ranging from 0% to -17%.
- It is recommened to repeat the live-load test using a few different load paths to get more robust rating estimates. Using multiple load paths and the resulting average load ratings, the discrepancy of the estimates is reduced to 2-11% for the four bridges.

- The results from sensitivity analysis of model updating-based methods do not clearly indicate that any of the sensing configurations are more appropriate than the others, but do illustrate that comparable results can be achieved using a more limited sensor suite. The results suggest that the updating process is not constrained by the sensor configuration. However, a general principle within a St-Id framework is the assurance that the measurements used in a model updating strategy are inclusive of response characteristics that are activated during testing (e.g. deflection measurements at midspan during a live load test, vertical accelerations for flexural impact test, or support rotation measurements for a midspan loading).

**Acknowledgements**

The research team would like to acknowledge the support provided by VDOT in execution of this research. This acknowledgement includes the technical review panel members: Prasad Nallapaneni, Jonathan Mallard, John Lindemann, and Ed Hoppe, for providing guidance on the project direction as well as the various personnel from the Staunton and Richmond districts who were involved in the testing and evaluation programs.